\documentclass[12pt]{emulateapj}
\usepackage{natbib}

\slugcomment{AJ accepted}

\shorttitle{RCS2 survey construction}
\shortauthors{Gilbank et al.}

\begin{document}

\title{The Red-sequence Cluster Survey-2 (RCS-2): survey details and photometric catalog construction}

\author{David G.~Gilbank}
\affil{Department of Physics and Astronomy, University of Waterloo, Waterloo, Ontario,  N2L 3G1, Canada}
\affil{Department of Astronomy and Astrophysics, University of Toronto,, 50 St George Street, Toronto, Ontario, M5S 3H4, Canada}
\email{dgilbank@astro.uwaterloo.ca}

\author{M.~D.~Gladders}
\affil{Department of Astronomy and Astrophysics, 
University of Chicago, 5640 S. Ellis Ave., Chicago, IL, 60637, USA}

\author{H.~K.~C.~Yee}
\affil{Department of Astronomy and Astrophysics, University of Toronto,
  50 St George Street, Toronto, Ontario, M5S 3H4, Canada}

\and

\author{B.~C.~Hsieh}
\affil{Institute of Astrophysics and Astronomy, Academia Sinica, P.O. Box 23-141, Taipei 106, Taiwan}

\def\LaTeX{L\kern-.36em\raise.3ex\hbox{a}\kern-.15em
    T\kern-.1667em\lower.7ex\hbox{E}\kern-.125emX}

\def\logmass{$\log(M_*)$}
\def\ms{$\log(M_*/M_\odot)$}
\def\ha{{\rm H$\alpha$}}

\def\msunyr{M$_{\odot}$ yr$^{-1}$}

\def\kms{km s$^{-1}$}

\def\lsim{\mathrel{\hbox{\rlap{\hbox{\lower4pt\hbox{$\sim$}}}\hbox{$<$}}}}
\def\gsim{\mathrel{\hbox{\rlap{\hbox{\lower4pt\hbox{$\sim$}}}\hbox{$>$}}}}

\begin{abstract}
The second Red-sequence Cluster Survey (RCS-2) is a $\sim$1000 square degree, multi-color imaging survey using the square-degree imager, MegaCam, on the Canada-France-Hawaii Telescope (CFHT). It is designed to detect clusters of galaxies over the redshift range $0.1\lsim z \lsim$1. The primary aim is to build a statistically complete, large ($\sim10^4$) sample of clusters, covering a sufficiently long redshift baseline to be able to place constraints on cosmological parameters via the evolution of the cluster mass function. Other main science goals include building a large sample of high surface brightness, strongly gravitationally-lensed arcs associated with these clusters, and an unprecedented sample of several tens of thousands of galaxy clusters and groups, spanning a large range of halo mass, with which to study the properties and evolution of their member galaxies. 

This paper describes the design of the survey and the methodology for acquiring, reducing and calibrating the data for the production of high-precision photometric catalogs. We describe the method for calibrating our $griz$ imaging data using the colors of the stellar locus and overlapping Two-Micron All-Sky Survey (2MASS) photometry. This yields an absolute accuracy of $<0.03$ mag on any color and $\approx$0.05 mag in the $r$-band magnitude, verified with respect to the Sloan Digital Sky Survey (SDSS). Our astrometric calibration is accurate to $\ll 0.3$\arcsec~from comparison with SDSS positions. RCS-2 reaches average 5$\sigma$ point source limiting magnitudes of $griz = [24.4, 24.3, 23.7, 22.8]$, approximately 1-2 magnitudes deeper than the SDSS. Due to the queue-scheduled nature of the observations, the data are highly uniform and taken in excellent seeing, mostly FWHM$\lsim$0.7\arcsec~in the $r$-band.  In addition to the main science goals just described, these data form the basis for a number of other planned and ongoing projects (including the WiggleZ survey), making RCS-2 an important next-generation imaging survey. 
\end{abstract}

\section{Introduction}
\label{sec:introduction}

Clusters of galaxies are ideal tracers of the largest density fluctuations in the Universe, and their abundance (and its evolution with cosmic time) may be used to place constraints on cosmology (e.g., \citealt{ecf96}). They also provide ideal laboratories for studying galaxy evolution. Originally used in this way since they contain a large number of galaxies all in the same location, it has since become clear that the properties of their member galaxies are markedly different from galaxies in the general field (e.g., \citealt{dressler80}, \citealt{1999ApJ...527...54B}, \citealt{Ellingson:2001zo}), implying that mechanisms which truncate star formation and transform galaxy morphology operate on cluster scales (e.g., \citealt{Treu:2003ty} and references therein). 

Constructing large, well-defined samples of galaxy clusters has a long and varied history. The first systematic searches involved visual identification of overdensities of optical galaxies on photographic plates \citep{abell,aco}. In the 1970s, with the advent of X-ray telescopes above the Earth's atmosphere, selection of clusters from their extended X-ray emission found favour \citep{mitchel76,serlemitsos77}. Recently, a combination of large format CCD detectors and objective algorithms to search efficiently for signatures of galaxy clusters has led to a revival in the use of optical selection in cluster surveys \citep{pdcs,kep,gal,gy00,2004MNRAS.348..551G}. A variety of techniques have been suggested to exploit the expected luminosity and/or color distribution of galaxies in clusters. The big advantage of these surveys compared with the older visual searches is that the detection method could be automated and characterised, meaning that the survey selection function could be quantified. Arguably the most efficient method is that of \citet{gy00} which uses the fact that the cores of galaxy clusters are dominated by galaxies with old stellar populations, forming a tight red-sequence in color magnitude space \citep{visv, bow92}. A number of other realisations of red-sequence based cluster finding algorithms exist (e.g., \citealt{Koester:2007ek}) differing in some details but all relying on accurate colors from imaging in two or more filters. The observed color of this sequence provides an accurate distance estimate. The application of this method led to the construction of the first Red-sequence Cluster Survey (RCS-1, \citealt{Gladders:2005oi}), a 72 square degree imaging survey in two bands ($R_C$ and $z^\prime$) designed to locate galaxy clusters from $0.2\lsim z \lsim1.1$ using the technique of \citet{gy00}. 

Not only is optical selection of galaxy clusters undergoing a revival, but astronomy in general is entering an era of `survey science' where an unprecedented number of wide-field optical (and NIR) surveys are currently underway or planned, such as LSST \citep{LSST-Science-Collaborations:2009uq}; Pan-STARRS\footnote{see http://pan-starrs.ifa.hawaii.edu/public/}; UKIDSS \citep{Lawrence:2007rp}; and DES\footnote{see https://www.darkenergysurvey.org/}. In addition, many of these wide-field optical surveys are specifically targeted at areas surveyed for clusters using other methods, such as the Blanco Cosmology Survey \citep{High:2010qy} of the South Pole Telescope (SPT, \citealt{Carlstrom:2009ys}) Sunyaev Zel'dovich effect (SZ)-selected cluster survey. For surveys using other methods (such as SZ selection), the optical data are critical for the verification of the cluster candidates found and for the determination of photometric redshifts. Furthermore, surveying the same areas with multiple techniques allows important comparisons of the different selection effects and the resulting properties of the clusters found (e.g., \citealt{don01,2004MNRAS.348..551G,Rasmussen:2006qy}).

In this paper we describe the second Red-sequence Cluster Survey (RCS-2), the largest survey of this new generation for which imaging has already been completed. This builds on the methodology of RCS-1. The RCS collaboration has invested a large amount of work in attempting to characterize the selection function and the properties of clusters selected with this technique. Many of these results are directly applicable to RCS-2 (such as mass--richness calibrations) and so it is useful to summarize some of the RCS work to date. 

The efficiency of the selection method employed by the red-sequence surveys is that it can locate and estimate the redshifts of clusters using only one color (two filter) data, given the appropriate choice of filters. It is impractical to obtain mass estimates from follow-up observations of the $\sim$30 000 clusters which will be found in RCS-2, so the survey data themselves must be used to produce a proxy for cluster  mass. Significant, representative samples of clusters from RCS-1 have been followed up using a variety of mass estimators such as dynamical \citep{gilbank:07a,felipe07}, X-ray \citep{hicks07}, strong and weak-lensing (from an ACS snapshot programme, PI:Loh; ACS SNe Cosmology project PI:Perlmutter) and SZ observations. In this way, the relationship between our mass proxy (optical richness from the survey data) and cluster mass can be understood. 

One of the primary goals of RCS-1 was to place constraints on cosmological parameters ($\Omega_M$, $\sigma_8$, \citealt{Gladders:2007us}) via the growth of the cluster mass function. This demonstrated for the first time the feasibility of such an approach using an optically-selected cluster sample. This approach used the measured relation between mass and richness, but also showed that meaningful constraints could be obtained using a self-calibration technique \citep{Majumdar:2004nj} to estimate the form of this relation from the survey data themselves.  These authors demonstrate that the best constraints are obtained when accurate mass estimates are available for a subsample of clusters within the survey. It is worth emphasising that even if there is significant scatter in the relation between mass and the proxy (as we have found for optical richness), it is only important that the size of the scatter be well understood. With an order-of-magnitude larger survey than RCS-1, it becomes feasible to also constrain the equation of state of dark energy, $w$, \citep{Majumdar:2004nj} and this is in part the motivation for RCS-2. RCS-1 also produced a significant sample of strongly gravitationally lensed arcs around the clusters found. The number and redshift distribution of these lensing clusters were used to argue about the physical properties of the clusters responsible for their lensing cross-section and the relevance of such systems to constraining cosmology \citep{Gladders:2003xo}. The identification of such high surface brightness, strongly-lensed galaxies is another primary science driver for RCS-2. The massive clusters can be used as gravitational telescopes for studying high-redshift galaxies \citep[e.g.,][]{Pettini:2000fk,Wuyts:2010ul} which would otherwise be too faint to observe in detail. The giant arcs can also be used as probes of the properties of the cluster lenses themselves.

With a statistical sample of galaxy clusters, such as in RCS-1, it is possible to study the properties of their member galaxies (e.g., their luminosity functions and blue fractions) by stacking subsamples built from the survey data themselves \citep{Gilbank:2007rq,Loh:2007be}. With the order-of-magnitude larger RCS-2, it becomes feasible to measure much weaker trends and push measurements of cluster galaxies down to much lower overdensities. The addition of photometric redshifts (e.g., \citealt{Hsieh:2005fq}) will allow these techniques to be extended to the field environment. Such galaxy evolution studies will be explored in future work with RCS-2.

The outline for this paper is as follows. In \S2 we give an overview of the survey design and observations; \S\S3 \& 4 deal with CCD pre-processing, reduction, object detection and photometry; \S5 describes the photometric calibration via accurate fits to the star colors in our survey fields; \S6 outlines the procedure for and accuracy of the astrometric calibration. \S\S7 \& 8 describe the incorporation of additional data into our primary RCS-2 catalogs: $i$-band data which covers a large ($\sim$70\%) subsample of the primary $g$, $r$, $z$ survey area; and public imaging data from the CFHTLS-Wide survey, respectively. \S9 describes the final cleaning of the photometric catalogs: stitching into contiguous patches, removing duplicate data between overlapping pointings, and masking of artefacts. \S10 summarises and describes ongoing and future work for the survey. 

\section{Survey Overview}

The wide-field imaging capability of the square degree imager, MegaCam \citep{Boulade:2003uk}, on the 3.6-m CFHT makes it feasible to carry out a survey covering a significant fraction of the sky in a modest amount of observing time. Coupled with the exquisite seeing conditions attainable from the summit of Mauna Kea, such a survey can achieve impressive depths/resolution within this time. 

The main science drivers for RCS-2 dictate a survey area of $\sim$1000 deg$^2$ with the ability to detect galaxy clusters out to z$\sim$1 using the method of \citet{gy00}. Following the strategy of RCS-1 \citep{Gladders:2005oi}, which used $R_C$- and $z$-band imaging, $r^\prime$- and $z^\prime$-band\footnote{Hereafter we use the shorthand of omitting the prime notation from the MegaCam filter names.} filters are chosen to cover the bulk of this redshift range. In order to better distinguish lower-redshift clusters, $g$-band is added to accurately measure the colors of galaxies at z$\lsim$0.4, once the 4000\AA~break begins to move blueward of the $r$-filter. The $g$-band filter also helps to identify strongly gravitationally-lensed arcs around clusters, since the former tend to be relatively blue. Thus, the survey comprises three-color, $g$, $r$, $z$, imaging over the whole area. $i$-band imaging is obtained for the majority of this area via a data-exchange with the Canada-France High-z Quasar Survey \citep{Willott:2005eu}.

\subsection{Observations}

\begin{figure}
	{\centering
	\epsscale{1.3}
	\plotone{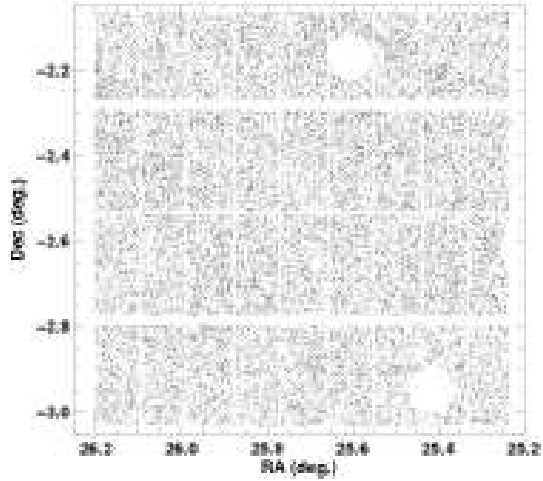}
	\caption{Example of a catalog produced for one MegaCam pointing, illustrating the layout of the CCDs. Points show $r<24$ objects classified as galaxies in one example pointing (0133A0). The 9$\times$ 4 grid of the individual detectors is clearly visible, as are the large gaps between the top and bottom rows and the others, described in the text. Circular regions absent of galaxies show where our masks for bright star haloes (described in \S\ref{sec:masking}) have been applied.}
	 \label{fig:ccds}
	}
\end{figure}

MegaCam comprises 36 CCDs arranged in a four row $\times$ nine column grid at the prime focus of CFHT. Each chip is 2048$\times$4612 pixels with a pixel scale of 0.187\arcsec,  allowing it to adequately sample the exquisite seeing possible from Mauna Kea. The spacing between CCDs is approximately 13\arcsec, with larger gaps ($\approx$80\arcsec) between the uppermost and lowermost rows and the other CCDs. This gives sky coverage of 0.96$\times$0.94 deg$^2$ (see Fig.~\ref{fig:ccds}). 

Observations were carried out in queue-scheduled mode on CFHT on runs between semesters 2003A and 2007B inclusive. PI imaging time was granted through requests to Canadian and Taiwanese agencies. 

Exposure times were set to 4, 8, and 6 minutes in $g$, $r$, and $z$ respectively. In 0.65\arcsec~seeing, according to the MegaCam exposure time calculator, these should correspond to  5-$\sigma$ point source limits of $g\approx$25.3, $r\approx$24.8, and $z\approx$22.5. The depth was set in the two reddest bands ($r$ and $z$) by the requirement to reach $\approx M^\star+1$ red-sequence cluster galaxies at z$\sim$1. Part way through our survey observations, it was found that 0.65\arcsec~seeing was not available on as many nights as required by our program per semester\footnote{This was somewhat mitigated by the significant improvement in Megacam image quality resulting from the L3 lens being replaced in the upside-down position in late 2004.}. So, a two-tier strategy for $r$ and $z$-band image quality was adopted in which only half the survey would be conducted in the 0.55-0.75\arcsec~bracket and half would be conducted in 0.75-0.90\arcsec. This means that the lowest richness clusters will not be detected in the worse seeing imaging (due to the reduced depth) all the way out to z$\sim$1, but richer clusters will. Since the former are exponentially more numerous, this should have negligible impact on cosmological constraints. $g$-band imaging was always performed in the better-seeing bracket in order to preserve the low surface brightness requirement for the detection of strongly-lensed arcs. Our final measured depths are detailed in \S\ref{sec:magcal}.

Single exposures were taken at each pointing position, i.e., no dithering was used. This is because sufficient depth is achieved in only a single exposure of 4-8 minutes (depending on filter), so the overhead associated with reading out the CCD ($\approx$2 minutes) becomes a significant fraction of the integration time if this is divided into two or more exposures. The priority is to cover as large an area of sky as possible to these depths  in a given amount of observing time. With this strategy the survey will contain chip gaps, but these can be dealt with (for cluster finding, etc.) via geometric correction factors (discussed in \S\ref{sec:stitching}). Gaps in the data also appear due to the occasional failure of a CCD within the mosaic camera.\footnote{In a small number of cases, problems with the camera electronics have caused half the mosaic to fail and so the queue observers have observed the northern and southern half of a single pointing in two separate observations with MegaCam operating in this half functional mode. In these cases we have processed the `half pointings' as if they were separate pointings with an 18 chip camera and combined them into the final catalog at the stitching stage (\S\ref{sec:stitching}).} 

\begin{figure*}
	{\centering
	\epsscale{1.}
	\plotone{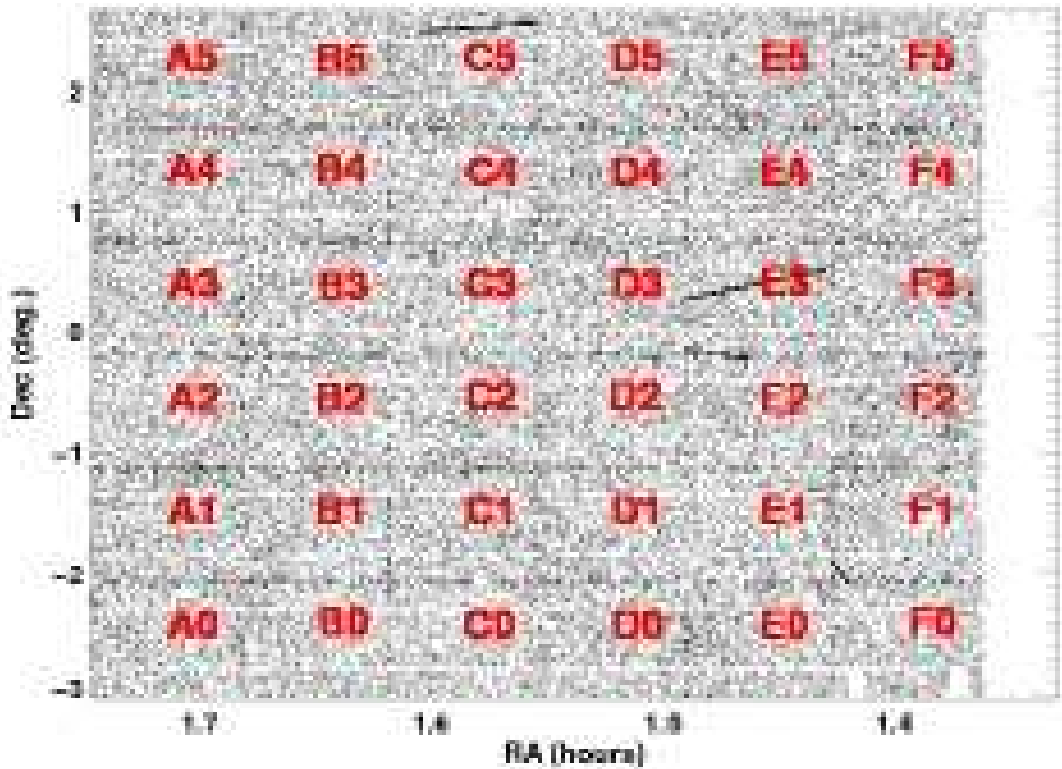}
	\caption{Layout of a typical patch (0133). Points show objects in the photometric catalog brighter than $r<20$. The chip gaps within each pointing are clearly visible. Pointings are labelled with their two character names, as described in the text. Overdensities of objects are visible where pointings overlap, as these duplicate areas have not yet been removed in the stitching process (see \S\ref{sec:stitching}). Artefacts such as satellite trails have not yet been cleaned (see \S\ref{sec:masking}) and examples of these can be seen in pointings C5 and E3. Additional gaps are visible from where one of the CCDs (lower right corner) was non-functional when E0 and F0 were observed.}
	 \label{fig:patch}
	}
\end{figure*}

Individual pointings are arranged in a grid pattern in discrete patches/fields (discussed in the next section). Pointings are named with the patch name, followed by a letter designating the column, beginning at 'A' running east to west. Next a digit specifies a pointing's location in the declination direction, beginning at '0', running from south to north. Thus, pointings within patch '0133' are labelled '0133A0' to '0133F5', from the south-east to north-west corners across the patch. Each pointing overlaps with its neighbors by $\approx$1 arcmin, typically. See Fig.~\ref{fig:patch}. 

The original observing strategy was to observe a given pointing sequentially in all three filters ($g$, $r$, $z$), but the overheads associated with filter changing made this relatively inefficient and so, after the first semester, the queue observers switched to observing several neighboring pointings in a single filter and then repeating these in the next filter. $i$-band imaging is typically performed many months later for a given pointing. This is potentially a consideration for projects wishing to use RCS-2 for time variability studies.

\subsection{Field descriptions}

When conducting a large survey, it is obviously advantageous to place fields such that they overlap with other large surveys at other wavelengths, whenever possible. The RCS-2 survey area is divided into discrete patches (where `patch' refers to a contiguous set of MegaCam pointings). The patches are divided approximately into two categories: those designed to overlap with other surveys (such as SWIRE, \citealt{Lonsdale:2003ix}) are typically a $6 \times 6$ grid (i.e., 6 degrees on a side); and those designed to be `blank' pointings typically $9 \times 9$. The larger patches are designed to be sufficiently large that the clustering of galaxy clusters may be used to provide additional constraints on cosmology (e.g., \citealt{Majumdar:2004nj}). These larger `blank' pointings are placed in the footprint of the UKIDSS \citep{Lawrence:2007rp} wherever possible, so that we will obtain NIR data for our optical imaging, useful for deriving stellar masses of galaxies, for example. The patch size is reduced to $\sim6 \times 6$ where we target other surveys, if the field size of the survey with which we overlap is smaller than $6\times6$ deg$^2$. We choose discrete patches distributed across the sky in order to improve observing efficiency (by distributing fields in RA), and to minimise the effects of cosmic variance (although this would only be an issue for the rarest systems, such as the most massive clusters, since our individual patches are so large - the largest patches in RCS-2 are already bigger than the entire RCS-1!). Most of our fields are equatorial in order to maximize follow-up from telescopes in both hemispheres. 

\begin{figure*}
	{\centering
	\plotone{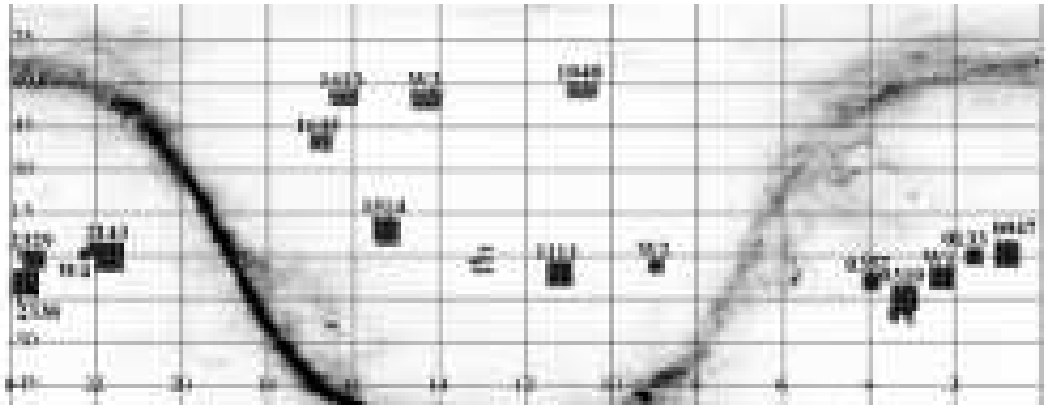}
	\caption{Layout of RCS-2 patches on the sky in Cartesian projection, plotted as a function of RA (hours) and Dec (degrees). Grey-scale shows distribution of Galactic Extinction from \citet{1998ApJ...500..525S}. Squares denote individual Megacam pointings. (Horizontal gaps between pointings in high declination patches are an artefact of the projection and the pointings do in fact overlap, as described in the text.) Patches are labelled with patch names. Additional (unlabelled) pointings around 13 hours equatorial denote a patch which was left uncompleted (1303) and will not be considered as part of the survey proper (due to the non-contiguous nature of the observed regions), but individual pointings will still be included in searches for strong lenses. }
	 \label{fig:map}
	}
\end{figure*}

All patches are chosen to be in regions of low Galactic extinction (see Fig.~\ref{fig:map}). Another consideration is the avoidance of bright stars. Using the Bright Star Catalogue \citep{Hoffleit:1964rt}, patches are shifted slightly to minimise the number of stars they contain in the range $2 < m_V < 4$ wherever possible. Stars this bright (and indeed somewhat fainter) can cause large reflection artefacts (see \S\ref{sec:masking}) which can render significant areas of the image unusable. Stars brighter than $m_V<2$ are potentially hazardous to the detectors and so pointings in our grid which would overlap with such stars are not observed.

A list of the patches observed is given in Table~\ref{tab:patches}. Patch 1040 covers the Lockman Hole, also surveyed by SWIRE and UKIDSS-DEX (Deep Extragalactic survey), as well as partially covered by UKIDSS LAS; 1613 also covers a SWIRE (ELAIS-N1) and UKIDSS-DEX field; 1645 covers a SWIRE (ELAIS-N2) field. 2329 targets a DEEP2 field. In addition to these targeting specific surveys, our survey fields have been used by other surveys. The WiggleZ survey \citep{Drinkwater:2010zm} has surveyed several of our patches (0047, 0133, 2143, 2329, 2338) with GALEX in order to select z$\gsim$0.5 Lyman Break Galaxy candidates using optical/UV colors to generate a large spectroscopic redshift sample to measure cosmological parameters via the signature of baryon acoustic oscillations imprinted on their clustering. The Canada France High-z Quasar Survey (CFHQS, \citealt{Willott:2005eu}) uses our $z$-band imaging in conjunction with their own $i$-band imaging to search for high redshift (z$\sim$7) quasars.  

These fields total 785 one-square degree pointings of imaging data for RCS-2. To this we add the four patches of the publicly-available Wide component of the CFHLS which tallies 171 deg$^2$ in $u^\star$, $g^\prime$, $r^\prime$, $i^\prime$, $z^\prime$. The reformatting and recalibration of these data to resemble the reduction of our PI imaging data is described in \S\ref{sec:cfhls}. Thus, overall, RCS-2 comprises 955 one-square degree pointings.

\begin{deluxetable}{lcccp{3.5cm}}
\tablecolumns{5}
  \tabletypesize{\small} \tablecaption{RCS-2 primary survey (PI imaging)  \label{tab:patches}}
  \tablewidth{0pt} \tablehead{ \colhead{Patch\tablenotemark{a}} & \colhead{R.A.} &
    \colhead{Dec.\tablenotemark{b}} & \colhead{extent\tablenotemark{c}} &
    \colhead{comments\tablenotemark{d}}
  } \startdata
\cutinhead{RCS-2 primary survey (PI imaging)}
0047 & 00:47.4 & +00:45 & 9$\times$9 & \\
0133 & 01:33.2 & -00:10 & 6$\times$6 & \\
0310 & 03:10.3 & -14:11 & 9$\times$9 & Initial patch extended to better cover the WMAP cold spot\\
0357 & 03:57.2 & -08:48 & 6$\times$6 & \\
1040 & 10:40.9 & +57:48 & 6$\times$6 & Lockman Hole\\
1111 & 11:12.0 & -05:52 & 9$\times$9 & \\
1514 & 15:14.6 & +08:55 & 9$\times$10 & \\
1613 & 16:13.6 & +55:00 & 6$\times$6 & ELAIS-N1 \\
1645 & 16:45.9 & +39:35 & 6$\times$6 & ELAIS-N2 \\
2143 & 21:41.0 & -00:07 & 10$\times$10 & \\
2329 & 23:26.1 & -01:14 & 8$\times$6 & DEEP2 \\
2338 & 23:38.5 & -09:07 & 9$\times$9 & \\
\cutinhead{CFHT Legacy Survey Wide component}
W1 & 02:18.0 & -07:27 & 9$\times$8 & XMM-LSS\\
W2 & 08:57.8 & -03:18 & 5$\times$5 & \\
W3 & 14:17.9 & +54:30 & 6$\times$7 & \\
W4 & 22:11.4 & +01:48 & 6$\times$6 & VVDS \\
\enddata
\tablenotetext{a}{patch name}
\tablenotetext{b}{RA \& Dec (J2000) of the patch centre}
\tablenotetext{c}{size (in number of 1 deg$^2$ pointings).This gives the approximate extent of the patch, but the layout is not necessarily rectangular. The exact geometry can be seen in Fig.~\ref{fig:map}. \label{tab:patches}}
\tablenotetext{d}{other relevant comments, such as overlapping surveys}. 

\end{deluxetable}

\section{Pre-processing}
Standard CCD pre-processing (bias subtraction, bad-pixel masking, flatfielding, etc.) is performed by the  
Elixir Project\footnote{see {\tt http://www.cfht.hawaii.edu/Instruments/Elixir/}} at CFHT. For the interested reader, an excellent description of the raw MegaCam data, including example images at various stages of Elixir reduction, may be found at {\tt http://www.cfht.hawaii.edu/Science/CFHTLS-DATA/rawdata.html}. 

We use the default Elixir reduction to handle all these preliminary steps, except for the $z$-band defringing which we found could be improved upon. Due to the large volume of data generated by our observations, rather than starting from raw (non-defringed) data and attempting to generate new master fringe frames from other $z$-band observations taken at similar times to our observations\footnote{due to the queue-scheduled nature of the observations, it is not always possible to find sufficient $z$-band data taken within a short enough time-frame to do this anyway.}, we investigated the possibility of starting with the Elixir (non-optimally)-defringed $z$-band images. It was found that running these images  through defringing code developed by the CFHT Supernova Legacy Survey (SNLS) Team (kindly provided by A. Conley) resulted in improvement to the defringing for all $z$-band images.\footnote{Briefly, this code uses principal component analysis (PCA), taking a set of principal components (the eigenvectors) of the $z$-band fringe pattern from data taken on an early SNLS observing run (using MegaCam). It then attempts to find the amplitude of the fringe pattern which must be subtracted (the eigenvalues) in order to remove the pattern.}
 However, for a small fraction of the $z$-band data ($\sim$5\% of all our images), significant fringing is still present. The initial fringe amplitude in the raw data was at a level of $\sim$15\% of sky\footnote{see {\tt http://www.cfht.hawaii.edu/Science/CFHTLS-DATA/rawdata.html}}. This was typically reduced to better than 5\% of sky by the initial Elixir defringing and reduced to negligible levels ($\lsim$0.5\% of sky) by the SNLS code for $\gsim$95\% of all our data. For the remaining $\sim$5\% of our $z$-band images, the residual fringing was at worst $\sim$3\% of sky. Thus, the most serious impact to our photometric measurements from the residual fringing would be an additional photometric error of $\approx$0.03 mag in $\sim$5\% of the $z$-band photometry added in quadrature to the intrinsic Poissonian error due to photon statistics. Thus, this inflates the photometric errors for very bright objects where the intrinsic error is $\lsim$0.03 mag; but has negligible impact on the vast majority of our objects, which are fainter and thus have larger intrinsic errors. The one aspect of our catalog construction impacted by the presence of low-level residual fringing is object detection. The implications for this are discussed in \S\ref{sec:ppp}. 

We note that our final images are identical to those generated by Elixir and stored in the CFHT archive (except for the $z$-band which have been further de-fringed). However, for each image, we also generate an associated bad pixel mask (BPM) using the procedures outlined below, which stores information about pixels contaminated with reflection haloes from bright stars, satellite trails, and cosmic ray hits, etc.

\subsection{Image alignment} 
\label{sec:ali}
An initial estimate of the world coordinate system (WCS) is taken from the Elixir solution written in the image headers. In many cases this solution can be offset from the real pointing position by several arcseconds, and a significant fraction of these show a non-physical instrumental solution (such that one or more of the chips appear to be overlapping or displaced from their nominal position in the camera grid). So, we construct a corrected WCS by running fast object-finding, with a high threshold set to find only a subsample of the brightest objects, and pattern-match these against an astrometric reference catalog. We use a model for the global placement of the chips within the camera and solve for the locations of all objects within the pointing simultaneously. This alleviates the problem with displacement of individual chips, from the Elixir solution, which would otherwise cause problems when we attempt to measure offsets between the different filters for each pointing. This quick calibration follows a similar method to that which provides our final, accurate astrometric calibration, which is described in detail in \S\ref{sec:astrom}. For each pointing, a WCS is generated for each of the three filters, $g$, $r$, $z$, and this is used to find the approximate offset between the different filters.

\section{Object detection and photometry}
\label{sec:ppp}
Object finding and photometry are performed using an updated version of Picture Processing Package ({\sc PPP}; \citealt{1991PASP..103..396Y}). While the basic methods are the same as those described in \citet{1991PASP..103..396Y} and \citet{Yee:1998rt}, because of the large amount of imaging data involved, a considerable number of additional algorithms have been implemented to allow the procedures to be carried out as a completely automated pipeline.  In the following, we provide brief descriptions of the basic methodology and the added features that allow the automation of the process.

\subsection{Object finding}
In RCS-1, object detection was performed on coadded $R_C+z$ images in order to increase the overall depth of the observations and to ensure the inclusion of the very reddest objects which may have been missed if the $R_C$-band alone (the deeper band) had been used.  As a result of the low-level ($\lsim$3\%) residual fringing in a small fraction ($\lsim$5\%) of the MegaCam $z$ data, it is not possible to everywhere use co-added $r+z$ images for object detection. (The $z$-band fringing means that spurious objects are more likely to be detected if they sit near the peak of a fringe, whereas some real objects will be missed if they sit in the valley of the fringe pattern. The higher noise in the sky caused by fringing also artificially increases the detection threshold, meaning that some genuine objects may be missed in clean areas.) So, for consistency, and to be conservative, object detection is performed solely on the $r$-band images. Our depths in $r$ and $z$ are matched (by construction) for z$\sim$1 red-sequence cluster galaxies, and so even though the catalog is formally $r$-selected it is also $z$-limited. Only for galaxies redder than z$\sim$1 cluster members will the catalog begin to become incomplete\footnote{this limit is tagged on a frame-by-frame basis by the {\sc PPP}-measured photometric limits.}. Therefore, the only slight limitation is that searching for extremely red objects ($r$-dropouts) is not possible with this catalog.

{\sc PPP} performs object detection on each chip as described in \citet{1991PASP..103..396Y}, with additional algorithms to automatically compute the detection threshold and to block out problem regions on the $r$-image. If artifacts (reflection haloes and saturated columns) associated with bright stars contaminate the chip being considered, the measurement of the sky noise level, which is used in the estimate of the object detection level, may be biased high. Furthermore, these regions will also contain many false object detections. So, a mask is applied to remove contaminated areas from consideration when estimating the sky level and the noise. 

To mask bright stars and their halos, the approximate WCS solution derived in \S3.1 (typically still accurate to better than 1\arcsec) is used to send a query to Vizier\footnote{see http://vizier.hia.nrc.ca/} and download a catalog of all bright stars from the Tycho 2 catalog \citep{Hog:2000ys} within the pointing. 
This bright star list is used to flag regions in which the sky is heavily contaminated by light from a bright star, and which must not be used when estimating the sky level for object detection. At this stage, a simple empirical relation between the observed size of the star halo and the cataloged magnitude of the star is used to set the mask size (which is set to be conservatively larger than might be needed). Later in the pipeline, a more detailed model of the properties of the reflection haloes around bright stars is used to delimit the area for object finding and to set mask flags in the final catalog (see \S\ref{sec:masking}).
The masking of saturated columns is done by identifying and connecting pixels above the saturation value starting from the positions of the bright stars. 
The masked areas are used neither in estimating
detection threshold, nor in object finding.

Object detection otherwise follows the same basic procedure as described in  \citet{Gladders:2005oi}. Peaks are identified as significant enhancements measured in a 3$\times$3 tapered box more than 2.6$\sigma$ above the local sky (excluding masked regions, as described above). 
Two criteria based on minimum number of connected pixels in a smoothed
image and the `sharpness' of the candidate in the unsmoothed image are used to eliminate cosmic ray and noise spike detections.

\subsection{Photometric measurement}
Total photometry of an object is derived based on the growth curve of
the object which is measured in a series of concentric circular apertures
around the object, masking nearby objects as required (see Yee 1991 for details). The growth curve is initially measured to 8.5\arcsec~for all objects. The center of the apertures is determined by an iterative procedure that is accurate to fractional pixels (down to better than 1/10 of a pixel, depending on the signal-to-noise ratio of the detection). To alleviate the problem of pixelation at small radii, each pixel is subdivided into 7$\times$7 subpixels before integrating the flux within an aperture. The photometric curve of growth so produced is used to identify an optimal aperture within which to measure the total magnitude following the procedure described in \citet{1991PASP..103..396Y}. 
If the optimal aperture of the object is 
smaller than the standard aperture of 8.5\arcsec~diameter, 
the magnitude within the optimal aperture is extrapolated to the standard 
aperture using corrections derived from the shape of the growth curves 
of bright point-source objects. 
The standard aperture size is chosen to include close to 100\%~of the light
for small objects under all our seeing conditions. For brighter resolved objects, which normally would still have increasing flux at the 8.5\arcsec~aperture, growth curves extended to a maximum diameter of 25\arcsec~are measured and used in determining the optimal aperture to make sure that the bulk of the light is included. We note that objects with an optimal aperture smaller than 1.5\arcsec~are classified as non-detections. The error on the total magnitude is then calculated as the sky noise within the photometric aperture used, which is sky noise-limited for the faint galaxies of interest here.

The object positions detected from the $r$ image are taken as the `master' position list. To perform photometry on the images of the other filters, the position list has to be transformed to the pixel coordinates of these images to an accuracy of about 2 pixels, so that {\sc PPP} is able to determine without ambiguity the subpixel centroids of the objects. For images that are offset from the parent $r$ image by less than 10 pixels, or, images whose offset can be estimated to an accuracy of better than 10 pixels from their approximate WCS solution
(\S3.1), a simple and very fast algorithm is used to register the position file. Here, 35 to 70 brightest non-saturated objects are identified in the $r$ image using a quick-find algorithm and used as position reference objects. Local maxima in the daughter frame within 10 pixels of the expected positions of the reference objects are located and used to derive average shifts in $x$ and $y$. A rotation, if one exists and is sufficiently small ($<\sim$1 degree),  is also measured by comparing vector angles between pairs of objects. This transformation is performed on the position list, and photometry
carried out on the daughter frame.

Occasionally, registration of the object positions for frames for the other filters of a pointing is not so straightforward due to a significant rotation and even a small scale change. This could occur when MegaCam is removed and re-installed, or the optics of the camera are changed (as in the case when one lens was inverted) or a filter is replaced (as in the case of the $i$-band filter, see \S\ref{sec:iband}). In these cases, more CPU-intensive algorithms are used to deal with the registration of the position files. The basic algorithm is a brute-force technique of finding the best $\chi$-square fit of the two sets of bright position reference objects, stepping through small increments of $xy$ shifts, rotation, and scale. The algorithm steps through increasingly more parameters in the above order, so as to minimize computational time. {\sc PPP} performs the overall registration process by starting with the simplest  procedure, tests the validity of the transform (based on whether the transformed positions from the $r$ frame match the actual objects on the other filter frames  close enough for {\sc PPP} to compute an accurate fractional pixel centroid), and if not, performs the next more complicated matching procedure. In most cases, it is the $i$-band images (discussed in \S\ref{sec:iband}), sometimes taken a year or more apart from the others, and comprising images taken through two different  filters over the whole survey, that require the more time-consuming registration procedures.

With growth curves for each object measured in all filters, colors are measured using a smaller aperture to improve the signal-to-noise ratio by minimizing the sky noise. The ``color aperture" used is either a 3\arcsec~diameter aperture or the optimal aperture, whichever is smaller.  The same aperture is used for each filter, taking the r-band as the reference.  No correction is made for seeing differences between the bands since the seeing values are typically comparable and the aperture is already relatively large compared to the PSF.  The total magnitude for each of the other filters is then calculated by  adding the relevant color to the total magnitude estimated in the reference filter ($r$).  The color error is then the quadrature sum of the errors in the two color apertures used to compute the color. Both color errors and total magnitude errors are propagated through to the final photometric catalogs. 

\begin{figure}
	{\centering
	\plotone{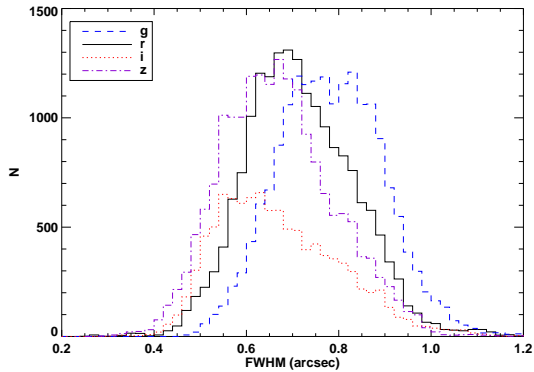}
	\caption{The distribution of seeing values (one value for each chip) in each filter. }
	 \label{fig:see}
	}
\end{figure}

A 5-$\sigma$ reference limiting magnitude for each chip is estimated by scaling the flux of a set of bright reference stars until they reach a signal-to-noise level of five. Typically for extended sources, the 100\% completeness limit is about 0.6 to 0.8 magnitudes brighter than this limit \citep{1991PASP..103..396Y}. An estimate of the seeing is made by measuring the average FWHM of the reference PSF stars identified in each chip.  The object catalogs typically comprise around 4000 objects per CCD chip. The values of seeing measured by {\sc PPP}, one value per chip, are shown in Fig.~\ref{fig:see}. The median values of the seeing are $[g,r,i,z] = [0.79\arcsec, 0.71\arcsec, 0.53\arcsec, 0.67\arcsec]$, but note that the distributions are broad due to our two-tier strategy.  A significant  fraction of our data possesses seeing better than $\lsim0.6$\arcsec in $r$.

\subsection{Star-Galaxy Classification}
Star-galaxy separation is performed by comparing the shape of the growth curve of each object to the weighted average growth curve from a set of four to eight reference PSFs from nearby unsaturated stars (see \citealt{1991PASP..103..396Y} and \citealt{Yee:1996lk}) chosen from a list of automatically identified reference PSFs from the same chip. The list of identified reference PSFs is created from all bright unsaturated objects that do not have close neighbors and  are deemed `not fuzzy' by an algorithm that iteratively eliminates objects that are `fuzzier' than the average. Each object is then given a classification of 0--artifact/spurious object ; 1 or 2--galaxy; 3--star; 4--saturated \citep{1991PASP..103..396Y}. The subdivisions of types 1 and 2 for galaxies is for historical reasons related to the development of PPP and no distinction is made in practice. An object is classified as saturated if one or more pixels is above the saturation ADU value as provide in the Elixir FITS header of the image. 

Based on tests using simulated images, the star-galaxy classification
is robust up to the 100\% completeness limit under typical seeing conditions \citep{1991PASP..103..396Y}.
At magnitudes fainter than 5$\sigma$, star-galaxy classification is
performed on a statistical basis using the ``variable classifier"
procedure based on the relative sharpness (rather than absolute
 sharpness) of the objects as described in \citet{1991PASP..103..396Y}.

\section{Photometric calibration}
\label{sec:photcal}
Accurate photometric calibration is one of the most important aspects of any imaging survey. For RCS-2, given the color-selected nature of our galaxy cluster sample, obtaining the highest possible accuracy in the calibration of each {\it color} is particularly important. Here we describe the technique we employ to calibrate the colors of galaxies as accurately as possible using the colors of stars in our fields.  

The photometric catalogs generated by {\sc PPP} all assume a single photometric zeropoint for all of the observations made in each of the three filters. Some of our observations were made in mildly non-photometric conditions, a condition we allowed after initial tests with our color-calibration technique demonstrated that we could achieve accurate calibration without the use of standard stars. Only mildly non-photometric conditions were allowed so as to not unduly decrease the depth of the observations, but this still increases the number of usable nights available to our project, since we do not require photometric conditions. 

In RCS-1, the $(R_C-z^\prime)$ color calibration of each pointing was performed by requiring the colors of bright galaxies to agree with a reference pointing. The $R_C$-band zeropoint was then set by requiring the number counts of galaxies to agree from pointing-to-pointing. In principal, the uniformity of the colors of stars or galaxies may be used. However, since we are primarily interested in extragalactic studies with RCS-2, we find it preferable to force-fit the properties of {\it stars} to agree from pointing-to-pointing in our survey, decoupling any calibration errors from the properties of the galaxies we wish to study. For example, forcing the number counts of galaxies to agree from field to field then makes it impossible to study density variations of galaxies on scales comparable to the size of each field, since we would have, by construction, forced all these values to agree. Thus in RCS-2 we will use the colors of stars to accurately set the calibration of each {\it color}, and then calibrate the value of the magnitudes in one reference band by comparison with an overlapping reference survey (2MASS).

\subsection{Fitting the stellar locus}

We use the uniformity of the colors of Galactic stars in order to set the color calibration from pointing-to-pointing in RCS-2. This method is an extension of the idea used by \citet{Hsieh:2005fq} to calibrate multicolor ($B$, $V$) follow-up observations of RCS-1 fields. \citet{Hsieh:2005fq} constructed histograms of objects classified as stellar and required that their $(z^\prime-R_C)$, $(V-R_C)$ and $(B-R_C$) distributions agree from pointing-to-pointing (holding the $R_C$-band calibration fixed). 

Our method is inspired by \citet{Ivezic:2004ad} who used the uniformity of the stellar locus in high-Galactic latitude fields in SDSS to assess the accuracy of their (independent) photometric calibration (using standard stars). They showed that, for high latitude ($|b|>15$ degrees) fields, the colors of faint stars are essentially identical and are dominated by stars in the Galactic halo (and thus behind the majority of the Milky Way dust). They showed that one can define principal colors along the stellar locus, and that constructing a histogram along these principal colors results in a Gaussian distribution where the width is dominated by the intrinsic width of the locus ($\sim$0.02 mag) if the photometric errors are small. Thus we combine the idea of principal stellar colors for checking the accuracy of photometry \citep{Ivezic:2004ad} with the color histogram-matching technique \citep{Hsieh:2005fq} to form a method for calibrating imaging surveys from the principal colors of stars. 

Our procedure is as follows. We begin by constructing a reference sample. Since our primary aim is a highly uniform color calibration across our entire survey, we design a calibration set using only MegaCam data to calibrate our filters internally to the native MegaCam system. We can then transform our (uniformly-calibrated) catalogs to any other (standard) system afterwards. To build a reference set, we select a subsample of around 20 RCS-2 pointings which overlap with SDSS. Both datasets are corrected for Galactic Reddening using the map of \citep{1998ApJ...500..525S},  since the majority of the ($|b|>15$ degrees) stars lie behind the Milky Way dust \citep{Ivezic:2004ad}. (All our fields lie at $|b|>30$ degrees, with the exception of the CFHT Legacy W2 field which is $|b|>20$ degrees.) For each pointing we match individual objects with their counterparts in SDSS and work out the median offset from SDSS for that pointing and filter. For example, for the $i^{th}$ pointing in the $r$-band filter, the natively calibrated magnitude, $r_{i, CAL}$, is related to the instrumental magnitude produced by PPP, $r_{i, instr}$, by the offset $dr_i$ by 
\begin{equation}
r_{i, CAL} = r_{i, instr} + dr_i
\end{equation}
where 
\begin{equation}
dr_{i} = median(r_{ij,SDSS}-r_{ij, inst})
\label{eqn:mediansdss}
\end{equation}
and $r_{ij,SDSS}$ and $r_{ij, inst}$ refer to the $j^{th}$ object matched between SDSS and MegaCam respectively. Now, for the $r$-band, the color term between MegaCam $r$ and SDSS $r$ is negligible, so this transformation effectively transforms onto the SDSS AB system. For the other filters there are color terms. Adopting the above procedure for all filters (i.e., neglecting to specifically apply a color term) means that our reference set is calibrated onto a native MegaCam system which agrees with SDSS for objects with the colors of the median-color objects in SDSS. We derive the color terms which may be applied to convert the RCS-2 native system to SDSS AB at the end of this section.

\begin{figure}
	{\centering
	\plotone{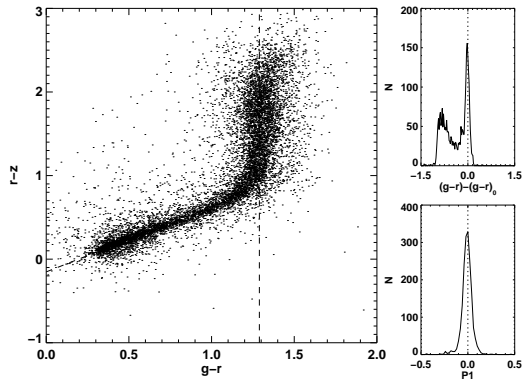}
	\caption{Main panel: color--color diagram of a subsample of stars used in the reference dataset. Dashed lines indicate the positions of the principal colors used. The vertical line is the $(g-r)$ principal color which is simply $(g-r)-(g-r_0)$ where $(g-r)_0$ is the reference color. A histogram of this principal color is shown in the upper right panel. The lower right panel shows a histogram of the P1 principal color, described in the text and indicated as the sloping dashed line in the main panel. See text for details.}
	 \label{fig:starcols}
	}
\end{figure}

We can now take each (uncalibrated) science pointing in turn and calibrate it to agree with the above 
 set. We do this for every pointing, including those which were used in constructing the reference sample, to ensure complete consistency across the survey. For calibrating $g$, $r$, $z$ data we will calibrate the $(g-r)$ and $(r-z)$ colors, initially holding the $r$-band magnitude fixed. We start with the $(g-r)$ color since the red branch of the stellar locus is almost independent of $(r-z)$ color (Fig.~\ref{fig:starcols}, see also fig. 2 of \citealt{Ivezic:2004ad}).  

The $(g-r)$ principal color is obtained by simply subtracting the $(g-r)$ color of the peak of the $(g-r)$ historgam, which we denote $(g-r)_0$, where $(g-r)_0=1.289$. The resulting histogram, now centred on zero color, is shown as the upper right panel in Fig.~\ref{fig:starcols}. 

The principal color used to calibrate $(r-z)$ is defined by fitting a linear relation to the blue part of the color--color diagram (Fig.~\ref{fig:starcols}):
\begin{equation}
P1 = (r-z) - [0.779(g-r) -0.152 ] \cap (g-r)<1.158.
\label{eqn:p1}
\end{equation}
The $(g-r)$ color cut ensures that only the blue branch of the stellar locus is used. This results in a very clean Gaussian in the histogram, shown as the lower right panel in Fig.~\ref{fig:starcols}. 

\subsection{Application of the fitting procedure}
With the above reference set in-hand, each science pointing may be examined in turn and its star colors forced to agree with that of our reference set. All objects classified as unsaturated stars with magnitude errors $<0.2$ mag are extracted and color histograms built in the same way as for the reference sample. The $(g-r$) histogram in the science pointing is cross-correlated with the of the reference pointing, and the offset $\Delta(g-r)_1$ found. This procedure is typically found to be accurate to a few hundredths of a magnitude ($\approx0.05$), by comparing the result with an independent calibration such as SDSS. In order to improve this accuracy further, a Gaussian profile is fitted iteratively around the position of the peak of the histogram, and the peak position and width recorded. The offset between the Gaussian-fitted peak and zero is a refined estimate of the color offset, $\Delta(g-r)_2$. The width of the fitted Gaussian, $\sigma_{(g-r)}$, is used to check for problems with the photometry/calibration method. The width of the distribution is set by the intrinsic width of the stellar locus convolved with our photometric errors, and thus should be approximately constant from pointing-to-pointing. If $\sigma_{(g-r)}>0.15$ mag then the pointing is flagged as having potential problems with the photometry. The inflated width of the Gaussian usually means that the $(g-r)$ colors are smeared out due to the  the $g$ and $r$ frames not having been registered correctly (occasionally due to the wrong pointing position being observed by the queue observers). A similar warning flag can be generated by flagging unusually low amplitudes of the Gaussian peak (since the number of stars in each pointing is approximately constant). Such calibration problems are later inspected manually. 

The calibration in $(g-r)$ is thus
\begin{equation}
(g-r)_{CAL} = (g-r)_{inst} + \Delta(g-r)_1 + \Delta(g-r)_2,
\label{eqn:dgr}
\end{equation}
where $ \Delta(g-r)_1$ is the offset determined by cross-correlation of the color histograms, and  $\Delta(g-r)_2$ is the offset from the peak of the iteratively-fitted Gaussian. In the case of a Gaussian width $\sigma_{(g-r)}>0.15$ mag, $\Delta(g-r)_2$ is set to zero and  a warning flag set. In a small minority of cases, this warning flag is caused by some problem other than bad registration of the images (such as contamination by satellite trails). In this case, if visual inspection of the calibration histograms reveals no obvious problem, then the calibration is accepted using just the cross-correlation offset, $\Delta(g-r)_1$. 

A similar procedure is applied in $(r-z)$, but using the principal color P1 (Eqn.~\ref{eqn:p1}) when constructing the  color histogram. Note that the $(g-r)$ color cut used with P1 to select the blue branch of the stars gives a very sharp, clean histogram in $(r-z)$. This can be done accurately since the $(g-r)$ color has been calibrated in the previous step. Again, a cross-correlation offset followed by a Gaussian fit to the located peak is performed and the same check, $\sigma_{P1}<0.15$ mag, made of the results.

\subsection{Magnitude calibration}
\label{sec:magcal}

\begin{figure}
	{\centering
	\plotone{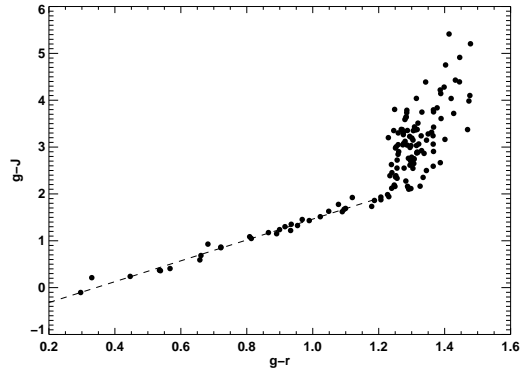}
	\caption{The reference pointing used to measure the $(g-J)--(g-r)$ locus by combining 2MASS and RCS-2 photometry. The blue branch in $(g-r)$ exhibits the smallest scatter and so this is fitted with the function in Eqn.~\ref{eqn:gJ} (dashed line). }
	 \label{fig:gJ}
	}
\end{figure}

The above procedure results in very accurate calibration of the $(g-r)$ and $(r-z)$ colors ($\lsim$0.03 mag). However, the  magnitude in any individual filter has not yet been accurately calibrated and may be considerably in error by several tenths of a magnitude (the data may have been observed through thin cloud, and even if not, we have made no explicit adjustment for atmospheric extinction). Operationally we held the $r$-band fixed above, but in order to calibrate the individual magnitudes it is sufficient to calibrate any one of the individual filters, provided we hold the colors fixed to their calibrated values. To obtain a calibration, we need a survey which overlaps with objects in our sample and possesses well-calibrated photometry. SDSS would be ideal, but it does not overlap with all of our pointings (or even all of our patches). So, we use NIR data from 2MASS\footnote{We thank our WiggleZ collaborators for originally suggesting this idea.}. $J$-band total magnitudes from the 2MASS point source catalog are downloaded for objects in the RCS-2 footprint. A reference set is created again using the SDSS-calibrated pointings, as above. Thus, we can construct color--color diagrams expected for stars in any combination of our ($+$2MASS) observed filters. We use $(g-J)$ versus $(g-r)$ as shown in Fig.~\ref{fig:gJ}. As mentioned above, we only need to calibrate one filter. Although $z$-band is closer in wavelength to observed $J$-band, it does not matter which combination is used, since the fit is performed in color--color space and the locus is just as well defined in these colors.\footnote{The reason for choosing $g$ is that this is the last of the three filters on which PPP performs photometry, so using it as the calibration filter is useful for catching problems with the photometry -- if the last filter ($g$) fails, then all the photometry fails for that pointing.} 

For $(g-J)$ the principal color used is: 
\begin{equation}
P2 = (g-J) - [2.226(g-r) -0.764] \cap (g-r)\le1.2. 
\label{eqn:gJ}
\end{equation}

The offsets, calculated as above, result in a $g$-band shift, $\Delta g$, which must be applied to each filter in such a way as to preserve the calibrated colors; i.e., if we re-write the original color shift (Eqn.~\ref{eqn:dgr}) as shifts to the individual filter, defining the $r$-band shift as zero, $\Delta g = \Delta(g-r)$ and $\Delta r=0$. The calibration offset measured from 2MASS, $\Delta g_{2MASS}$, gives resulting overall corrections to the instrumental magnitudes of
\begin{equation}
\Delta g_{tot} = \Delta g + \Delta g_{2MASS}
\end{equation}
and
\begin{equation}
\Delta r_{tot} = \Delta r + \Delta g_{2MASS}
\end{equation}
to preserve the calibrated $(g-r)$ color, and where $\Delta r$ was above defined as zero.

Similarly, for $z$ the offset would be
\begin{equation}
\Delta z_{Tot} = \Delta z + \Delta g_{2MASS},
\end{equation}
since its calibration was performed in $(r-z)$ and $r$ has just had $\Delta g_{2MASS}$ added to it.

Just to reiterate the procedure, we take data, which may have been acquired in non-photometric conditions, and obtain an accurate photometric calibration simply by forcing the colors of stars to agree with each other from pointing-to-pointing (which ensures highly uniform color calibration) and obtain a magnitude calibration by forcing these colors to agree with a reference set including $J$-band magnitudes from 2MASS in one of the colors. This procedure negates any need for separate correction of observational effects such as exposure time differences and atmospheric extinction.   We note that the idea of using stars to provide or test a photometric calibration is not a new one (indeed, \citealt{High:2009zp} recently presented an implementation of a similar technique), but we believe we are the first to implement such a method as the sole means of calibrating a survey of this size.

It is worth emphasising that since the stars used in the stellar locus-fitting are behind the majority of the Galactic dust, and we have reddening-corrected our reference dataset, the magnitude offsets measured between instrumental magnitudes and the reference set will correct the data to a system already corrected for Galactic Extinction. We note that in practice we prefer to both reddening correct our instrumental science catalogs and our reference dataset before performing the calibration. Thus, the magnitude differences, $\Delta mag$ (e.g., Eqn.~\ref{eqn:dgr}), can be directly applied to the initial instrumental magnitude catalogs (i.e., not corrected for Galactic Extinction) and this corrects the catalog initially to observed rather than reddening-corrected magnitudes. We find this more useful for considering observational effects (such as checking image depth). We store sub-pixel interpolated versions of the \citet{1998ApJ...500..525S} dust maps (to prevent possible sharp discontinuities across our patches caused by the large pixel size of the former) matched to our survey regions as part of this process. These then allow straightforward reddening correction of the observed magnitudes afterwards.

A second useful note about our procedure is that it is the almost vertical nature of the red branch of the stellar locus in $(g-r)-(r-z)$ which greatly simplifies the initial calibration.

\begin{figure}
	{\centering
	\plotone{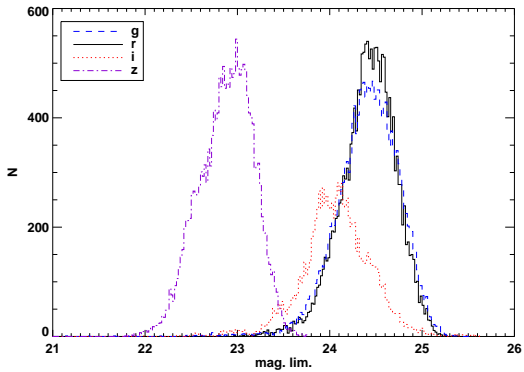}
	\caption{5$\sigma$ point source limiting magnitudes in each passband as measured by {\sc PPP}.}
	 \label{fig:lims}
	}
\end{figure}

The final depths of our imaging data are shown in Fig.~\ref{fig:lims}. Limits refer to the 5$\sigma$ point source limiting magnitude as measured by {\sc PPP}.

\subsection{Comparison of calibrated data with SDSS}
\label{sec:sdssphot}
In order to assess the accuracy of RCS-2 photometry, the significant overlap with the large, well-calibrated SDSS is used. In addition, this allows calculation of the transformation between the RCS2 MegaCam native system and the widely used SDSS system.

\begin{figure}
	{\centering
	\plotone{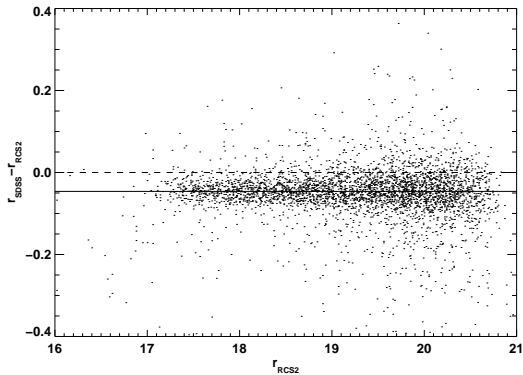}
	\caption{Object-by-object comparison of $r$-band total magnitudes from a single, randomly-selected RCS-2 pointing with objects in common with SDSS. Only objects unsaturated in RCS-2 are considered. The solid line shows the best fit to the $>$3600 objects in common. The dashed line shows zero offset for comparison. The offset is dominated by the calibration uncertainty in RCS-2, assuming a calibration error of $\Delta$m$\approx$0.01 mag in SDSS \citep{Ivezic:2004ad}. The scatter is dominated by photometric uncertainties in SDSS at the faint end, due to the shallower depth of SDSS relative to RCS-2. 
	}
	 \label{fig:dr_one}
	}
\end{figure}

The first step is to perform object-by-object matching for each pointing, as was done to construct the reference stellar locus dataset, as described in the previous section. To elucidate this, we show the comparison of a randomly selected RCS-2 pointing in Fig.~\ref{fig:dr_one}. This shows the comparison between $r$-band magnitudes in RCS-2 and SDSS for this pointing. The average offset between the two sets of calibrated magnitudes is estimated via a biweight fit to the difference which is shown by the solid horizontal line.  For this example pointing, the difference is 0.048 mag. The scatter is dominated by SDSS photometric errors at the faint end, due to the shallower SDSS photometry. At the bright end, the scatter is likely dominated by differences in the way magnitudes are measured by each survey. As described above, the RCS-2 magnitudes from {\sc PPP} are essentially estimates of the total magnitude, and these are compared with SDSS {\tt ModelMag} magnitudes. 

\begin{figure*}
	{\centering
	\plotone{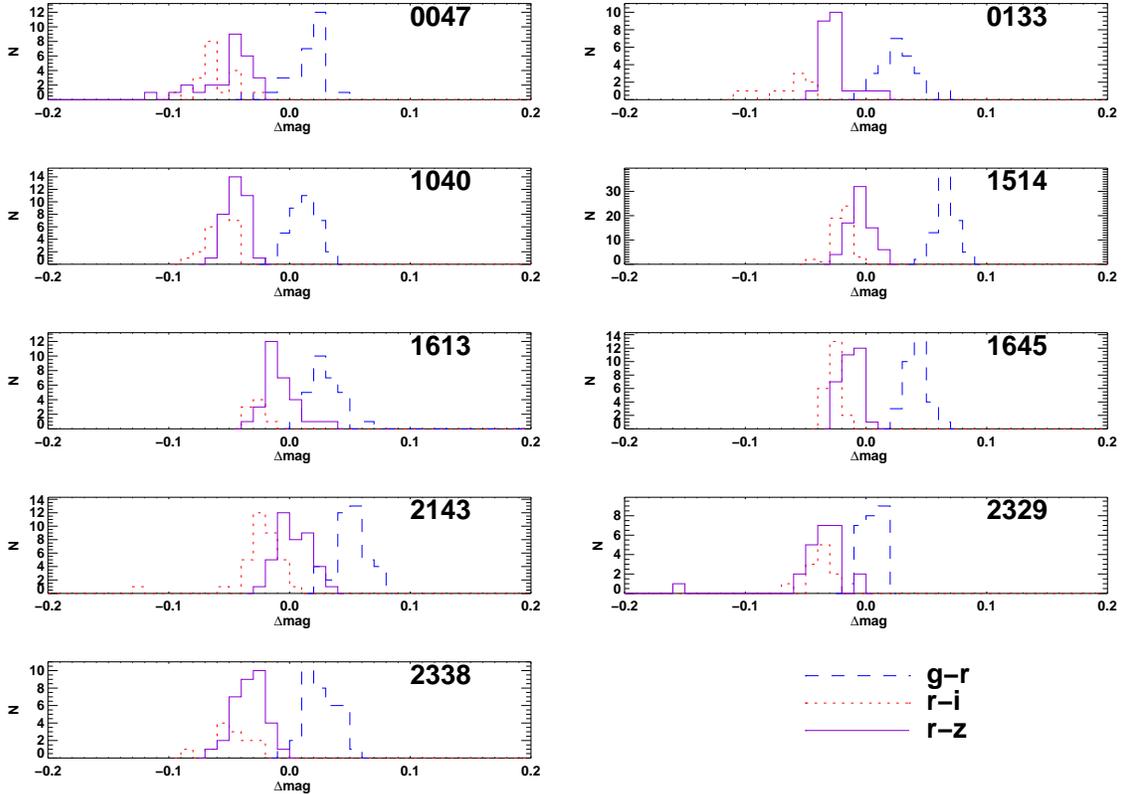}
	\caption{Check of the color differences from pointing-to-pointing by comparing RCS2 colors with those of SDSS. $\Delta mag=col_{RCS2}-col_{SDSS}$. Differences are measured object-by-object (as in Fig.~\ref{fig:dr_one}) and then a mean offset is assigned to each pointing. These histograms show the mean value of the offset for each pointing with significant SDSS overlap. Line styles denote different colors as annotated: $(g-r)$ dashed blue line, $(r-i)$ dotted red line, $(r-z)$ solid purple line. }
	 \label{fig:coldiff}
	}
\end{figure*}

Comparing colors relies on a more stable measurement than trying to estimate total magnitudes, so we begin here. Fig.~\ref{fig:coldiff} shows the differences in $(g-r)$ and $(r-z)$ and $(r-i)$\footnote{the $i$-band data are described in \S\ref{sec:iband}.} colors measured in RCS-2 compared with SDSS. $g$-band magnitudes have been corrected for the color term between MegaCam and SDSS $g$, as will be described below. Each panel shows a different RCS-2 patch and each value in the histogram represents the biweight average offset (as discussed above) difference between the two surveys for a single pointing. Only pointings with a significant number of objects (typically $>$1000) in common are considered. For each patch, the histogram in a given color is relatively narrow ($\lsim$0.05 mag), implying that the color calibration produces very uniform colors. There appear to be some small systematic offsets with respect to the SDSS photometry in the sense that the $(g-r)$ colors are systematically marginally redder than SDSS ($\sim$0.03 mag), whereas the $(r-i)$ and $(r-z)$ colors are slightly bluer ($\sim$0.04 mag). There also appears to be some evidence that the size of this small offset varies from patch to patch. This might be due to inaccuracies in the assumption that the Milky Way stellar populations used in the color calibration are identical for every patch (either due to intrinsic differences due to, e.g., metallicity of the stars, or inaccuracies in the assumption that they may all be considered behind all the Milky Way dust, as measured by the \citealt{1998ApJ...500..525S} dust map). Nevertheless, the calibration accuracy of the colors still represents a significant improvement over that usually achievable with the typical number of standard star observations made throughout an observing run. This is even more remarkable considering many of our data were taken in slightly non-photometric conditions. 

\begin{figure*}
	{\centering
	\plotone{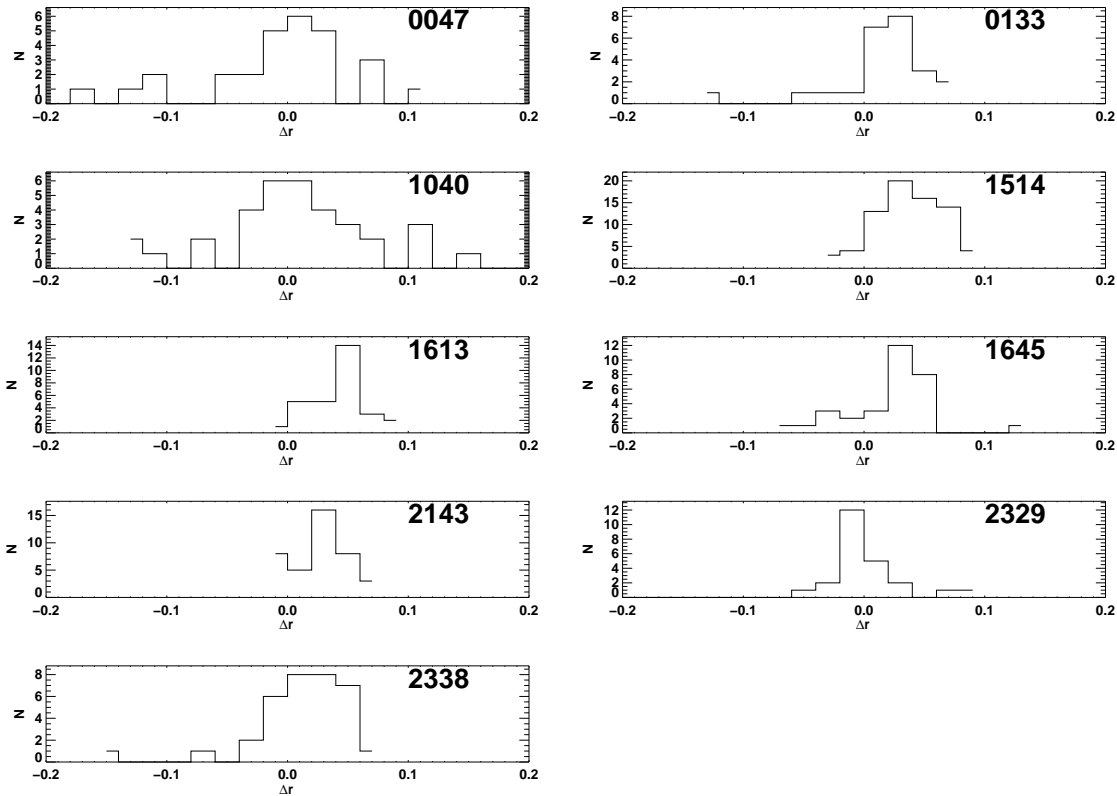}
	\caption{As for Fig.~\ref{fig:coldiff} but histograms now show the mean $r$-band magnitude offset for each pointing with respect to SDSS. }
	 \label{fig:rdiff}
	}
\end{figure*}

Fig.~\ref{fig:rdiff} is constructed in the same way as Fig.~\ref{fig:coldiff}, but the histograms now show the difference in $r$-band magnitude between RCS-2 and SDSS. The scatter is somewhat larger than the error in the color calibration, more like $\sim$0.05 mag rms, with some larger outliers. This likely reflects the limitation of using 2MASS photometry to set the magnitude zeropoint. It is difficult to find a sufficient number of common stars between 2MASS and typical 4-m telescope photometry, such that the stars are unsaturated in the larger telescope and sufficiently bright in 2MASS to yield a high S/N measurement of the star's magnitude. Even with the one square degree field of MegaCam, only $\approx$30 stars are typically suitable for use in the magnitude calibration.  

\subsubsection{Colour terms with respect to the SDSS system}

\begin{figure}
	{\centering
	\plotone{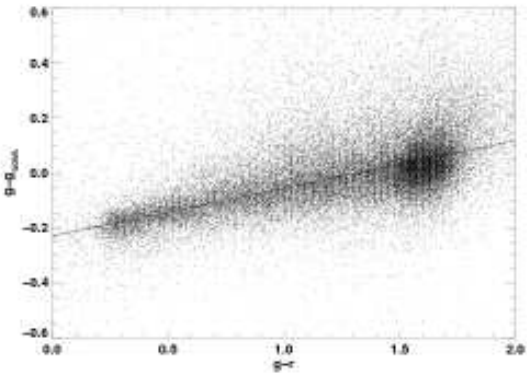}
	\caption{$g$ color term in $(g-r)$ deduced from a comparison with SDSS $g$ photometry. Plot shows MegaCam $g$ filter compared with SDSS, $g-g_{SDSS}$, as a function of MegaCam $(g-r)$ color for a whole RCS-2 patch. Solid line is color term equation given by Eqn.~\ref{eqn:gcolterm}.
	}
	 \label{fig:gcolterm}
	}
\end{figure}

Differences in the shape of the response of the MegaCam system (filter$+$CCD$+$optics$+$atmosphere) versus SDSS can be readily calculated following an extension of the above comparison. Individually-matched objects are again considered, but this time the average offset in each filter for each pointing is first subtracted (i.e., this would make the histograms in Fig.~\ref{fig:coldiff} and \ref{fig:rdiff} all zero-valued). We do this because we are only interested in the {\it shape} of the response, i.e., the slope of the magnitude difference as a function of MegaCam color, not its absolute value. Fig.~\ref{fig:gcolterm} shows this difference as a function of MegaCam $(g-r)$ color. All common, unsaturated objects ($>$40 000) in a single patch are used. As can clearly be seen, the $g$-band magnitude difference between the two surveys is a strong function of $(g-r)$ color. The best fit relation, shown as the solid line, is given by 
 \begin{equation}
 g_{SDSS} = g + 0.23 - 0.174(g-r).
 \label{eqn:gcolterm}
 \end{equation}
 As can be seen from Fig.~\ref{fig:gcolterm}, the best-fit line passes through zero offset from SDSS at the position of the median color [$(g-r)\sim1.6$], by construction, from Eqn.~\ref{eqn:mediansdss}. Due to the way in which we have internally color-calibrated RCS-2, this transformation to the SDSS system is now a trivial last step of the calibration procedure, which may be applied by using Eqn.~\ref{eqn:gcolterm} on the photometric catalog, if desired. 
 
\begin{figure*}
	{\centering
	\plottwo{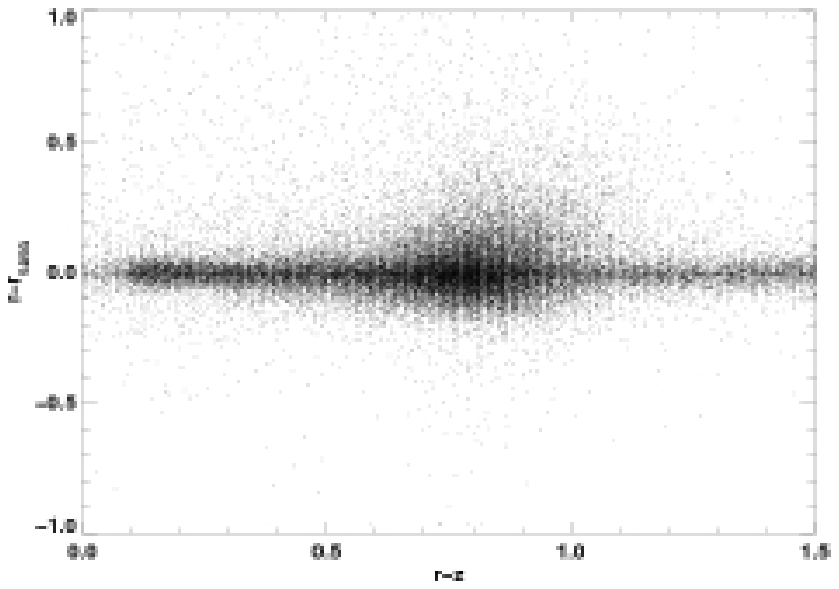}{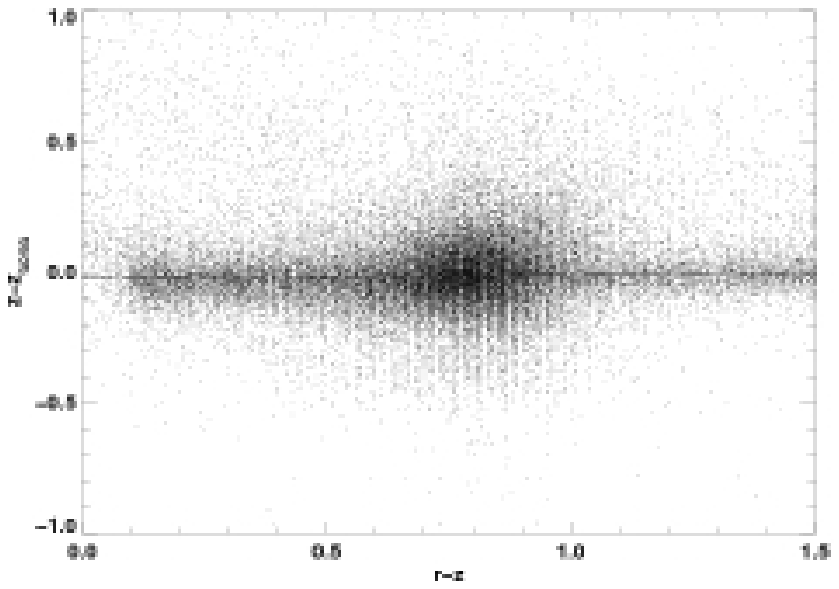}
	\caption{Estimates of $r$-band and $z$-band color terms with respect to the SDSS system from a comparison with SDSS photometry as a function of $(r-z)$ color for a whole RCS-2 patch. The solid line shows the best fit. For the $z$-band, the color term is 0.01 (which we ignore) and the $r$-band term is negligible.
	}
	 \label{fig:rzcolterms}
	}
\end{figure*}

Tests for color terms in $r$ and $z$ are performed in the same way. The results are shown in Fig.~\ref{fig:rzcolterms}. The $z$-band shows a tiny residual color dependence $\approx0.01(r-z)$ which would make $<$0.01 mag difference to the $z$-band data over the typical color range of objects and so may safely be neglected when comparing with SDSS photometry. The $r$-band shows no measurable color term. 

\subsection{Summary of photometric accuracy}

To summarise the systematic errors of our photometric catalogues, our color calibration technique is able to produce colors calibrated to to an absolute accuracy of $\lsim0.03$ mag. This is performed independently of the magnitude calibration (single filter) which is somewhat less accurate ($\lsim0.10$ mag) due to the need to tie to the 2MASS system, which has limited numbers of objects in each of our pointings. The 5-$\sigma$ point source depths of our catalogs (at which the magnitude error is $\approx$0.2 mag) are given in Fig.~\ref{fig:lims}. Our random magnitude errors are then limited by photon statistics in the usual way. One small exception to this is due to a small fraction ($\lsim$5\%) of our $z$-band data which possesses an additional error due to residual fringing at a level of $\lsim$0.03 mag. This makes a negligible additional contribution to the photometric errors for the vast majority of sources of interest to us for cluster detection (which are predominantly faint and have larger intrinsic photometric errors than this already). 

\begin{figure*}
	{\centering
	\plottwo{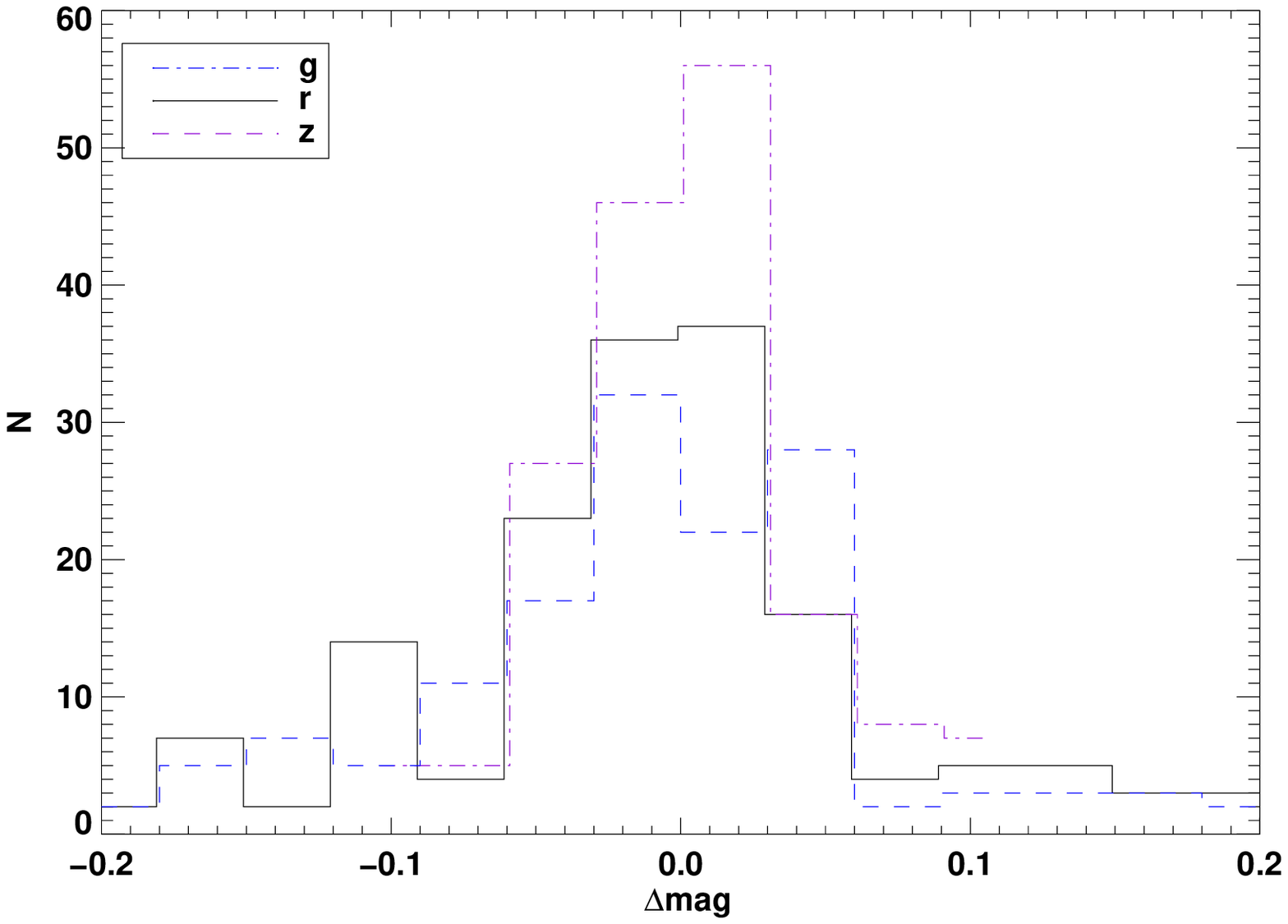}{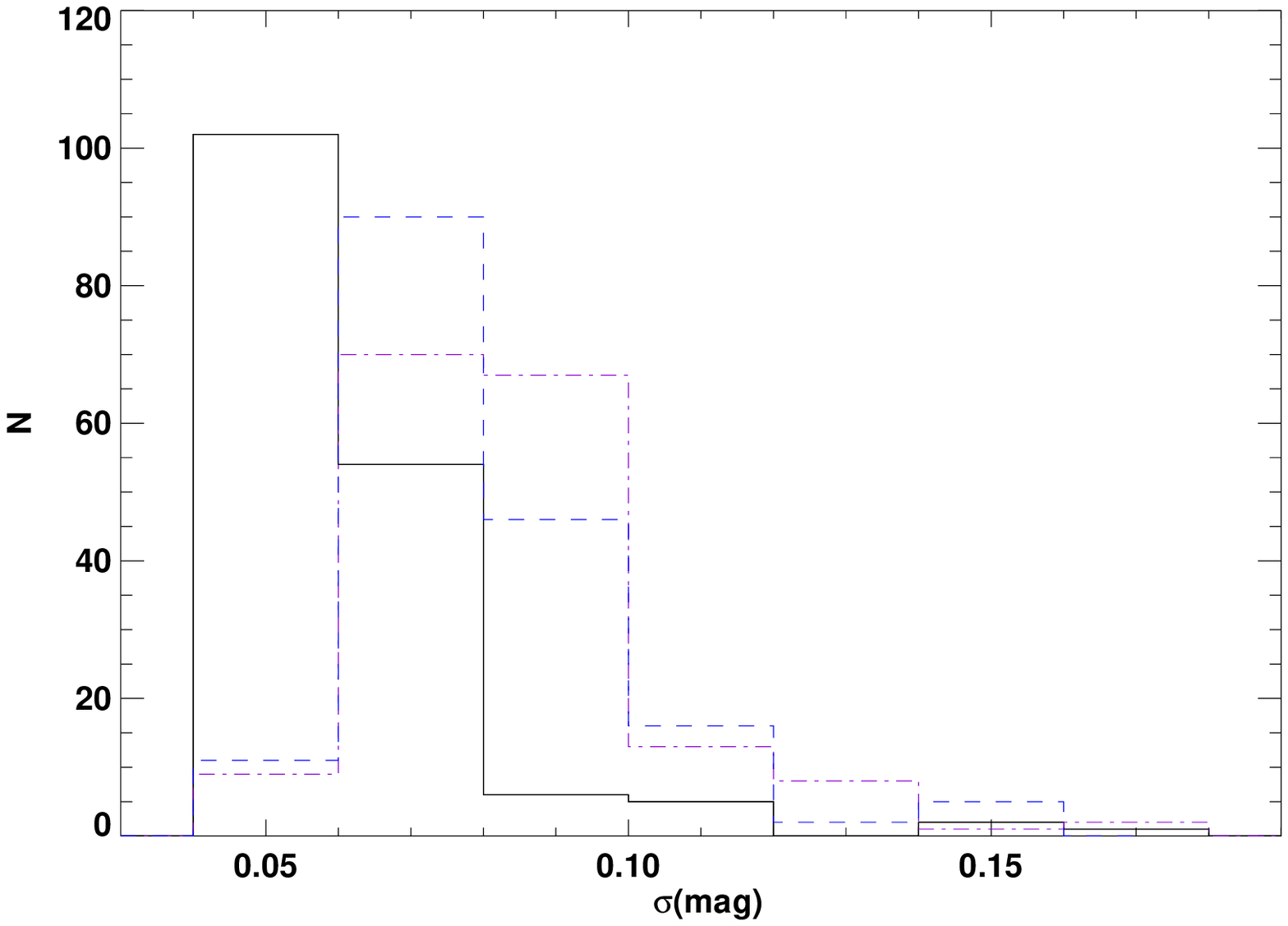}
	\caption{Internal accuracy of photometry estimated from all overlapping pointings within one patch (2143). The left panel shows the median difference for each pointing compared with its neighbor, and the right panel shows the median absolute deviation of these overlapping objects. Each entry in the histogram is the average of all points in the overlap region between a unique pair of pointings. Only overlaps containing more than 100 objects are considered. See text for discussion. 
		}
	 \label{fig:olaps}
	}
\end{figure*}

As one final test of the internal accuracy of our catalogs, we present a comparison of photometric differences between objects measured in the overlapping regions of neighboring pointings. Fig.~\ref{fig:olaps} shows the results for one representative patch, 2143. For each pointing, the difference between magnitudes in each of the $g$, $r$, and $z$ filters measured within that pointing and one of its neighbors is recorded. Only pairs of pointings with at least 100 objects in common are considered. The median difference between this pair of neighbors is then calculated and appears as a single entry in the histogram in the left panel of Fig.~\ref{fig:olaps}. The median absolute deviation is recorded for the same set of objects and is entered into the histogram in the right panel. The results are plotted for every unique pair of overlaps. The {\it rms} width of the difference histogram (left panel) is better than 0.04 mag, in good agreement with the estimates from the SDSS comparison. Similarly, the average intrinsic scatter within the overlap regions agrees with this. Tails toward greater values of the magnitude difference are seen, and illustrate one of the drawbacks of this comparison. The overlapping regions must necessarily come from the edges of the camera, where corrections for flatfielding, etc. are generally worst. Thus, comparing the edge in one pointing with a different edge of the camera in a neighboring pointing generally produces the worst agreement, and the internal accuracy of our catalogs are likely better than this over most of the field. Nevertheless, this is a useful, if somewhat pessimistic, estimate of the uncertainty. Similar results are seen for the other patches.

\section{Astrometric calibration}
\label{sec:astrom}
Astrometry is performed in a manner similar to that used in \citet{Gladders:2005oi}. An instrumental solution is first generated using an observation of an astrometric standard field, taken as part of the MegaCam calibration procedure. This is used to map individual chip coordinates to a global instrumental solution in terms of $\Delta$RA, $\Delta$Dec from the pointing centre. This is achieved semi-interactively using custom-written routines to fit a 2D cubic function using the IDL's {\sc polywarp} routine. In the absence of differential atmospheric refraction and other real-world limitations such as the reproducibility of mounting the camera on the telescope, this solution would be sufficient to describe the relative positions in every observation. In practice, it is necessary to fit for not only relative shift (of the pointing centre relative to the expected telescope pointing position) and rotation, but also for the differential displacement of objects across the field. This is achieved using lower-order corrections to the instrumental polynomial map, again using IDL's {\sc polywarp}. 

\begin{figure}
	{\centering
	\plotone{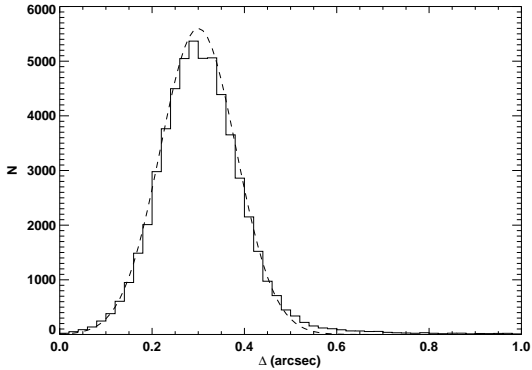}
	\caption{Astrometric offset between positions of SDSS objects and RCS-2 positions for all overlapping objects in a whole RCS-2 patch (2329). There are almost 60 000 galaxies in common. Dashed curve shows a Gaussian of $\mu=0.30$\arcsec and $\sigma=0.08$\arcsec.}
	 \label{fig:astrom}
	}
\end{figure}

For each pointing, x,y positions are taken from the PPP catalog. Now that the photometric calibration has been performed, it is a simple matter to extract a magnitude-selected sample in a similar way to an external astrometric reference catalog. The nominal object selection is $17 < r < 21$, which is matched against the full depth of the USNO-B1 astrometric catalog \citep{usno} using the instrumental solution as an initial estimate of the positions.\footnote{With a survey as large as RCS-2, even relatively rare occurrences can lead to a significant number of failures in automated routines. So, in practice we find it necessary to use a range of magnitude cuts if the nominal values fail to yield a sensible astrometric solution. In very rare cases (the presence of a rich globular cluster, M2,  and a nearby dwarf irregular galaxy, IC1613) it was necessary to intervene and reset parameters manually.} Shifts, rotation and distortion are found iteratively by proceeding in small steps. After each iteration, the {\it rms} offset from the USNO-B1 is measured and the results recorded after the final iteration. A comparison of object  positions from an entire RCS-2 patch with positions determined from SDSS is shown in Fig.~\ref{fig:astrom}. A generous matching radius of 3\arcsec~is used between the two catalogs to ensure that the difference is not underestimated due to distant matching objects being missed. The distribution of differences between object centres is shown with a Gaussian of mean, $\mu=0.30$\arcsec and width, $\sigma=0.08$\arcsec. The distribution has a higher mean offset than might be expected given its width (the mean and variance should be the same if the distribution is approximately Poissonian), and so the bulk of the offset may be due to systematic differences between the SDSS system and ours. A comparison between independent observations of the same objects in overlapping regions at the edges of our pointings shows that the rms difference is closer to 0.1\arcsec. Regardless, our astrometric precision is $\ll0.3$\arcsec, and this is comparable to (or better than) the accuracy of the photographic plates used in making the USNO-B1. This accuracy is confirmed by a large amount of spectroscopic follow-up with the wide-field spectrograph AAOmega by the WiggleZ collaboration \citep{Drinkwater:2010zm}, and by our own follow-up observations.

\section{Addition of $\MakeLowercase{i}$-band data}
\label{sec:iband}
\subsection{Image registration and photometry}

Although the RCS-2 survey proper comprises $g$, $r$, $z$ imaging, the majority of fields now also have $i$-band imaging obtained for the CFHQS \citep{Willott:2005eu} by reobserving our fields in $i$. We now discuss how these data have been incorporated, where available. The basic procedure is similar to that of the other filters, registering each $i$-band image to the completed $grz$ catalogs using the alignment method described in \S\ref{sec:ali}.  Once registered, photometry could proceed through {\sc ppp} as before. 

\subsection{Photometric calibration}
\label{sec:ical}
\begin{figure}
	{\centering
	\plotone{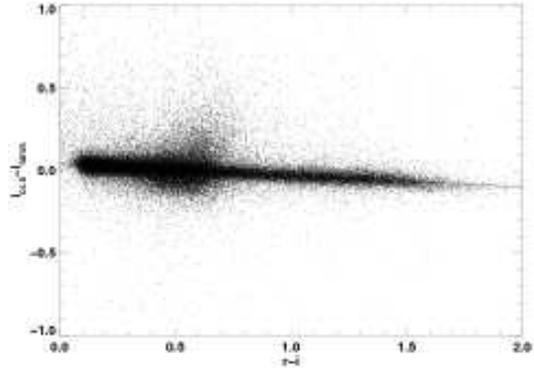}
	\caption{$i_{old}$ color term in $(r-i)$ deduced from a comparison with SDSS $i$ photometry. Plot shows MegaCam i.MP9701 filter compared with SDSS, $i_{old}-i_{SDSS}$, as a function of MegaCam $(r-i)$ color for a whole RCS-2 patch. Solid line is color term equation given by Eqn.~\ref{eqn:icolterm}.}
	 \label{fig:icolterm}
	}
\end{figure}

Due to the breakage of the original MegaCam $i$-band filter (i.MP9701) in June 2007, the survey contains $i$-band imaging from two different filters. The original filter was replaced with a new one (i.MP9702) possessing a much bluer response (close to the SDSS response). As a result, a significant color term exists between the two filters. For simplicity in the photometric catalogs, we decided to convert the i.MP9701 native photometry to that of the i.MP9702 native response. This has the advantage that the latter has a negligible color term compared with SDSS $i$, and thus our $i$-band data will be automatically calibrated to the AB system. The transformation between the two filters is found by comparing each with overlapping SDSS photometry using pointings in common, as described for the other filters in \S\ref{sec:sdssphot}. This is shown in Fig.~\ref{fig:icolterm}.

The conversion is
\begin{equation}
i_{new} = i_{old} + 0.067(r-i_{old}), 
\label{eqn:icolterm}
\end{equation}
where $i_{old}$ and $i_{new}$ refer to the i.MP9701 and i.MP9702 filters respectively. The rms scatter in this transformation is 0.034 mag.

\begin{figure}
	{\centering
	\plotone{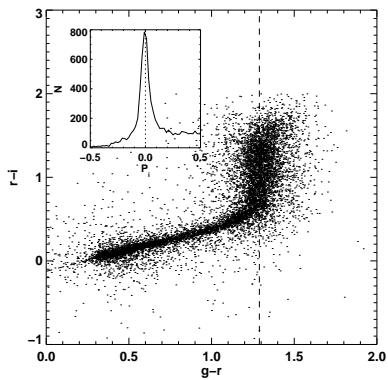}
	\caption{Colour--color diagram of the subsample of stars used as the reference set in the $i$-band calibration. Dashed lines indicate the principal colors used. The inset shows the histogram about the principal relation $P_i$ (Eqn.~\ref{eqn:p_i}), sloping dashed line.}
	 \label{fig:gr_ri}
	}
\end{figure}

Colour calibration of individual science pointings is achieved in the same way as for the $z$-band. Stars with magnitude errors $<$0.2 mag are selected from each pointing, and the stellar locus fitted to a reference locus in $(g-r)$--$(r-i)$, constructed as for the $(g-r)$--$(r-z)$ case. The principal color used (plotted in Fig.~\ref{fig:gr_ri}) is
\begin{equation}
P_i = (r-i) - [0.529(g-r)-0.125] \cap (g-r)<1.158
\label{eqn:p_i}
\end{equation}
Since the other bands have already been calibrated, a single offset to the $i$-band is all that is required.

\section{CFHTLS-Wide data}
\label{sec:cfhls}

The CFHT Legacy Survey\footnote{see {\tt http://www.cfht.hawaii.edu/Science/CFHLS/}} is a joint project between Canada and France to image a number of fields using MegaCam. The `Wide' component of the survey, considered here, covers 171 square degrees in $ugriz$. The observing strategy is somewhat different from RCS-2, since multiple exposures with dithering are used at each pointing position. Catalogues available to the Canadian and French communities (and later to the world) are produced by TERAPIX (Traitement Elementaire, Reduction et Analyse des PIXels de megacam)\footnote{see {\tt http://terapix.iap.fr}}. We use the T0006 internal release (2009 November) of the data. Briefly, the TERAPIX reduction involves similar pre-processing to the Elixir products, with the additional steps of determining astrometric solutions and combining the dithered images in each filter in each pointing into a stacked image. Then object detection and photometry is performed using {\sc SExtractor}  \citep{sex}. 

In order to make the TERAPIX catalogs resemble our method as closely as possible, we extract total magnitudes in the $r$-band using {\sc SExtractor}'s MAG\_AUTO estimate. We then construct total magnitudes in the $g$, $i$ and $z$ bands by constructing the relevant color with respect to $r$-band using aperture magnitudes measured in 4.4\arcsec diameter apertures and add these to the $r$-band total magnitude to mimic the method used by {\sc PPP}. i.e., 
\begin{equation}
g_{Tot} = (g-r)_{ap} + r_{Tot}
\end{equation}
for the $g$-band, where the subscripts $Tot$ and $ap$ denote total and aperture magnitudes, measured as described, respectively. Star/galaxy classification is  performed using {\sc SExtractor}'s CLASS\_STAR parameter. TERAPIX flag all objects brighter than $i<17.5$ as saturated regardless of whether they are actually saturated or not. To recover bright galaxies, which are not saturated, we follow the method of \citet{Lu:2009gf} considering the half-light radius estimated by {\sc SExtractor} as a function of $i_{Tot}$. 

The most important step in ensuring that the CFHTLS data and RCS-2 are on a uniform system is to ensure that the color calibrations are the same. $i$-band data are first converted from the old to the new filter system using the color term described in \S\ref{sec:ical}. 

The above procedures produce catalogs in which each pointing is independently calibrated in terms of photometry and astrometry. The final steps are to clean the catalogs, removing artefacts such as reflection halos of bright stars, satellite and meteor trails, etc., and to stitch the individual pointings in a patch into a contiguous area, removing duplicate objects in the overlaps between pointings (as described in \S\ref{sec:stitching}). The $g$, $r$, $i$, $z$ data are then run through the RCS-2 photometric calibration pipeline, as described in \S\S\ref{sec:photcal}, \ref{sec:ical}. Since the TERAPIX catalog construction already contains a photometric calibration, this typically results in small $\lsim0.05$ mag shifts in the zeropoint of each filter in each pointing. TERAPIX provide two alternate photometric calibrations for their catalogs. The default involves calibrating a single pointing in each of their patches and then using this as a reference. The calibration of all other pointings in the patch is then determined via the overlaps of all overlapping pointings, holding the single reference fixed. TERAPIX also provide photometric offsets which have been determined via Stellar Locus Regression using a technique similar, but not identical, to ours. In order to ensure complete uniformity across RCS-2, we use our stellar locus fitting and 2MASS comparison to calibrate the colors and magnitudes respectively, completely independently of the Elixir calibration. We find from comparison of each calibration with individual objects in SDSS (as in \S\ref{sec:sdssphot}) that our color calibration is somewhat more accurate than the TERAPIX color calibration ($\approx$0.02 mag vs $\approx$0.05 mag) but that our magnitude calibration is marginally worse ($\approx$0.05 mag vs $\approx$0.03 mag), again likely limited by the number of stars available from the overlap with 2MASS and their associated photometric errors (see Lu et al., in prep for a detailed comparison). It is possible that we will modify the final RCS-2 magnitude calibration to use internal overlaps as a final correction, but the current calibration is certainly sufficiently precise for our purposes.  

\section{Clean catalog construction}

\subsection{Artefact masking}
\label{sec:masking}

Two distinct sets of artefacts are present in the data which must be masked out to produce a clean photometric catalog. The first set is predictable artefacts such as those associated with bright stars. The positions of bright stars in MegaCam pointings are known a priori (once the astrometric calibration has been performed), and so the locations of halo reflection artefacts may be automatically predicted and masked. Secondary reflections within the camera lenses, or returned reflections off the detector or camera dewar window and then off an air-glass interface in the camera result in extended reflection haloes visible around bright stars. Careful examination of the images revealed three dominant reflection haloes. Each halo is of similar total intensity, and so the largest of these is visible above sky noise for only the brightest stars (typically at most a few stars per pointing) and the smallest is visible around relatively fainter stars (typically up to a few stars per mosaic chip). The outer boundaries of these artefacts are approximately circular, but the center of the reflection haloes of each type is a function of position within the camera reference frame. The halo sizes, and positions as a function of camera coordinates, were determined empirically from a set of test observations in each band, and these data were used to construct masks around known bright stars. The masking additionally requires a threshold brightness for each reflection halo; this was again determined empirically by examining object counts radially around bright stars, and choosing limiting magnitudes for each reflection halo that ensured that regions which are significantly compromised (either by having too many objects detected due to structure in the reflection halo, or by having too few objects due to the increased sky noise) are masked.  Objects in these masked regions are retained in the photometric catalog, but a flag is set warning that they are located in a reflection artefact and thus they may be straightforwardly rejected from subsequent analysis.  

The second set of artefacts are those whose locations cannot be predicted prior to examining the data, such as satellite trails and meteor trails (see Fig.~\ref{fig:patch}). These are dealt with in a semi-interactive manner. Since these are linear artefacts, line-finding techniques akin to the Hough transform (converting the positions of objects in the {\sc PPP} catalog to polar coordinates and looking for overdensities of points in this coordinate space) can be used to flag potential artefacts. In practice, the images are examined interactively with sites of potential artefacts flagged by this technique. At this point other artefacts which may have passed the automated masking are also flagged. These include occasional reflection haloes around stars whose as-observed magnitudes are significantly brighter than their cataloged values. Also, bright and very low redshift galaxies - typically subtending tens of arcseconds or more - will tend to produce many objects due to substructure in the galaxy. These regions are also masked to mitigate any influence of such objects on future analyses of object counts or clustering. Finally the most common masking added by hand at this stage is of diffraction spikes which have escaped the automated masking done by PPP (see \S\ref{sec:ppp}). Typically these are either from close but resolved double stars which appear as single entries in the catalogs used to initially mask the images and which produce two sets of diffraction spikes neither of which is caught by the initial mask, or diffraction spikes from stars which fall in the mosaic chip gaps. After all additional masking is complete, a mask image is constructed for each chip, with the values of the masked pixels set to indicate the source of the masking. These images also include a record of missing filter data along image edges (due to misalignments between images of differing filters) with missing data in each filter flagged by a different mask value. In principle, this allows one to select subsets of data masked only for certain reasons, while including other nominally masked objects; this may be of interest under some circumstances.

\subsection{Patch stitching and random catalogs}
\label{sec:stitching}

Up until this point, the data processing has been performed on individual pointings. Due to the survey strategy, overlaps occur between neighboring pointings. The final step is to `stitch' the overlapping pointings into single contiguous patches, removing duplicate objects which occur in the overlapping regions. This is performed in the same manner as for RCS-1. Briefly, the midpoint of the joint area between neighboring pointings is found and then each field is truncated at the midpoint. See \citet{Gladders:2005oi} for details. Thus, only objects in unique, non-overlapping areas are retained in the stitched photometric catalog. 

At the same time as this stitching procedure is being performed, catalogs containing a uniform surface density of points placed at semi-random positions (referred to as `random catalogs') are produced for the purposes of area calculation. These contain points placed across each patch at a nominal density of  0.1 arcsec$^{-2}$. Where these fall outside the survey geometry, or within regions masked by artefacts, the points are removed. Thus, these essentially represent a sparse `good pixel mask' for the survey. Future analyses requiring knowledge of the area considered around a given point (such as the area surveyed within X arcmin radius of a given galaxy cluster) can then count the number of points in the random catalog within that region to calculate the usable area. If higher density sampling is required, additional realisations at the same sampling density can be generated to increase the resolution as needed.

\section{Summary and future work}

We have presented the methodology for the reduction and precise calibration of the second Red-sequence Cluster Survey (RCS-2). The primary purpose of this survey is to build the largest sample of optically-selected galaxy clusters out to lookback times around half the age of the Universe. Results from cluster-finding (which will be presented in a forthcoming paper) show that the red-sequence redshifts are better than in RCS-1, due to the better uniformity and accuracy of the photometry in RCS-2, with typical $|z_{RCS}-z_{spec}| \ll 0.02$. A large number of projects aimed at characterizing clusters found with our technique have been ongoing for several years. An accurate understanding of the mass--richness relation will allow the cluster survey to place constraints on cosmological parameters. Future papers will present not only the cosmology results, but results on galaxy evolution, both within the clusters and the field. In addition to the large number of cluster science works possible, many other studies including those concerned with the properties of the Milky Way, such as the stellar populations and searches for dwarf galaxies, are possible. The large and accurate dataset provided by RCS-2 make it an important wide-field imaging survey of the new generation.

\acknowledgments

We are grateful to the Canadian and Taiwanese TACs for the generous allocations of CFHT time which made this project possible. We thank Alex Conley for providing the SNLS defringing code and for useful discussions on MegaCam data; Ben Koester for his assistance with masking bright stars and artefacts; and all our RCS and WiggleZ collaborators for thoroughly testing preliminary versions of these catalogs. 

The RCS2 project is supported in part by grants to HKCY from the Canada Research Chairs program, the Natural Science and Engineering Research Council of Canada, and the University of Toronto.

Based on observations obtained with MegaPrime/MegaCam, a joint project of CFHT and CEA/DAPNIA, at the Canada-France-Hawaii Telescope (CFHT) which is operated by the National Research Council (NRC) of Canada, the Institute National des Sciences de l'Univers of the Centre National de la Recherche Scientifique of France, and the University of Hawaii. This work is based in part on data products produced at TERAPIX and the Canadian Astronomy Data Centre as part of the Canada-France-Hawaii Telescope Legacy Survey, a collaborative project of NRC and CNRS.

\bibliographystyle{apj}

\begin{thebibliography}{50}
\expandafter\ifx\csname natexlab\endcsname\relax\def\natexlab#1{#1}\fi

\bibitem[{{Abell}(1958)}]{abell}
{Abell}, G.~O. 1958, \apjs, 3, 211+

\bibitem[{{Abell} {et~al.}(1989){Abell}, {Corwin}, \& {Olowin}}]{aco}
{Abell}, G.~O., {Corwin}, H.~G., \& {Olowin}, R.~P. 1989, \apjs, 70, 1

\bibitem[{{Balogh} {et~al.}(1999){Balogh}, {Morris}, {Yee}, {Carlberg}, \&
  {Ellingson}}]{1999ApJ...527...54B}
{Balogh}, M.~L., {Morris}, S.~L., {Yee}, H.~K.~C., {Carlberg}, R.~G., \&
  {Ellingson}, E. 1999, \apj, 527, 54

\bibitem[{{Bertin} \& {Arnouts}(1996)}]{sex}
{Bertin}, E. \& {Arnouts}, S. 1996, \aaps, 117, 393

\bibitem[{{Boulade} {et~al.}(2003){Boulade}, {Charlot}, {Abbon}, {Aune},
  {Borgeaud}, {Carton}, {Carty}, {Da Costa}, {et~al.}}]{Boulade:2003uk}
{Boulade}, O., {Charlot}, X., {Abbon}, P., {Aune}, S., {Borgeaud}, P.,
  {Carton}, P., {Carty}, M., {Da Costa}, J., {et~al.} 2003, in Presented at the
  Society of Photo-Optical Instrumentation Engineers (SPIE) Conference, Vol.
  4841, Society of Photo-Optical Instrumentation Engineers (SPIE) Conference
  Series, ed. {M.~Iye \& A.~F.~M.~Moorwood}, 72--81

\bibitem[{{Bower} {et~al.}(1992){Bower}, {Lucey}, \& {Ellis}}]{bow92}
{Bower}, R.~G., {Lucey}, J.~R., \& {Ellis}, R.~S. 1992, \mnras, 254, 601+

\bibitem[{{Carlstrom} {et~al.}(2009){Carlstrom}, {Ade}, {Aird}, {Benson},
  {Bleem}, {Busetti}, {Chang}, {Chauvin}, {et~al.}}]{Carlstrom:2009ys}
{Carlstrom}, J.~E., {Ade}, P.~A.~R., {Aird}, K.~A., {Benson}, B.~A., {Bleem},
  L.~E., {Busetti}, S., {Chang}, C.~L., {Chauvin}, E., {et~al.} 2009, arXiv
  0907.4445

\bibitem[{{Donahue} {et~al.}(2001){Donahue}, {Mack}, {Scharf}, {Lee},
  {Postman}, {Rosati}, {Dickinson}, {Voit}, {et~al.}}]{don01}
{Donahue}, M., {Mack}, J., {Scharf}, C., {Lee}, P., {Postman}, M., {Rosati},
  P., {Dickinson}, M., {Voit}, G.~M., {et~al.} 2001, \apjl, 552, L93

\bibitem[{{Dressler}(1980)}]{dressler80}
{Dressler}, A. 1980, \apjs, 42, 565

\bibitem[{{Drinkwater} {et~al.}(2010){Drinkwater}, {Jurek}, {Blake}, {Woods},
  {Pimbblet}, {Glazebrook}, {Sharp}, {Pracy}, {et~al.}}]{Drinkwater:2010zm}
{Drinkwater}, M.~J., {Jurek}, R.~J., {Blake}, C., {Woods}, D., {Pimbblet},
  K.~A., {Glazebrook}, K., {Sharp}, R., {Pracy}, M.~B., {et~al.} 2010, \mnras,
  401, 1429

\bibitem[{{Eke} {et~al.}(1996){Eke}, {Cole}, \& {Frenk}}]{ecf96}
{Eke}, V.~R., {Cole}, S., \& {Frenk}, C.~S. 1996, \mnras, 282, 263

\bibitem[{{Ellingson} {et~al.}(2001){Ellingson}, {Lin}, {Yee}, \&
  {Carlberg}}]{Ellingson:2001zo}
{Ellingson}, E., {Lin}, H., {Yee}, H.~K.~C., \& {Carlberg}, R.~G. 2001, \apj,
  547, 609

\bibitem[{{Gal} {et~al.}(2000){Gal}, {de Carvalho}, {Odewahn}, {Djorgovski}, \&
  {Margoniner}}]{gal}
{Gal}, R.~R., {de Carvalho}, R.~R., {Odewahn}, S.~C., {Djorgovski}, S.~G., \&
  {Margoniner}, V.~E. 2000, \aj, 119, 12

\bibitem[{{Gilbank} {et~al.}(2010){Gilbank}, {Barrientos}, {Gladders}, {Yee},
  {Ellingson}, {Infante}, {Hall}, {Hertling}, {et~al.}}]{felipe07}
{Gilbank}, D.~G., {Barrientos}, L.~F., {Gladders}, M.~D., {Yee}, H.~K.~C.,
  {Ellingson}, E., {Infante}, L., {Hall}, P.~B., {Hertling}, G., {et~al.} 2010,
  in prep

\bibitem[{{Gilbank} {et~al.}(2004){Gilbank}, {Bower}, {Castander}, \&
  {Ziegler}}]{2004MNRAS.348..551G}
{Gilbank}, D.~G., {Bower}, R.~G., {Castander}, F.~J., \& {Ziegler}, B.~L. 2004,
  \mnras, 348, 551

\bibitem[{{Gilbank} {et~al.}(2007){Gilbank}, {Yee}, {Ellingson}, {Gladders},
  {Barrientos}, \& {Blindert}}]{gilbank:07a}
{Gilbank}, D.~G., {Yee}, H.~K.~C., {Ellingson}, E., {Gladders}, M.~D.,
  {Barrientos}, L.~F., \& {Blindert}, K. 2007, \aj, 134, 282

\bibitem[{{Gilbank} {et~al.}(2008){Gilbank}, {Yee}, {Ellingson}, {Gladders},
  {Loh}, {Barrientos}, \& {Barkhouse}}]{Gilbank:2007rq}
{Gilbank}, D.~G., {Yee}, H.~K.~C., {Ellingson}, E., {Gladders}, M.~D., {Loh},
  Y.-S., {Barrientos}, L.~F., \& {Barkhouse}, W.~A. 2008, \apj, 673, 742

\bibitem[{{Gladders} {et~al.}(2003){Gladders}, {Hoekstra}, {Yee}, {Hall}, \&
  {Barrientos}}]{Gladders:2003xo}
{Gladders}, M.~D., {Hoekstra}, H., {Yee}, H.~K.~C., {Hall}, P.~B., \&
  {Barrientos}, L.~F. 2003, \apj, 593, 48

\bibitem[{{Gladders} \& {Yee}(2000)}]{gy00}
{Gladders}, M.~D. \& {Yee}, H.~K.~C. 2000, \aj, 120, 2148

\bibitem[{{Gladders} \& {Yee}(2005)}]{Gladders:2005oi}
---. 2005, \apjs, 157, 1

\bibitem[{{Gladders} {et~al.}(2007){Gladders}, {Yee}, {Majumdar}, {Barrientos},
  {Hoekstra}, {Hall}, \& {Infante}}]{Gladders:2007us}
{Gladders}, M.~D., {Yee}, H.~K.~C., {Majumdar}, S., {Barrientos}, L.~F.,
  {Hoekstra}, H., {Hall}, P.~B., \& {Infante}, L. 2007, \apj, 655, 128

\bibitem[{{Hicks} {et~al.}(2008){Hicks}, {Ellingson}, {Bautz}, {Cain},
  {Gilbank}, {Gladders}, {Hoekstra}, {Yee}, {et~al.}}]{hicks07}
{Hicks}, A.~K., {Ellingson}, E., {Bautz}, M., {Cain}, B., {Gilbank}, D.,
  {Gladders}, M.~D., {Hoekstra}, H., {Yee}, H.~K.~C., {et~al.} 2008,
  \apj, 680, 1022

\bibitem[{{High} {et~al.}(2010){High}, {Stalder}, {Song}, {Ade}, {Aird},
  {Allam}, {Armstrong}, {Barkhouse}, {et~al.}}]{High:2010qy}
{High}, F.~W., {Stalder}, B., {Song}, J., {Ade}, P.~A.~R., {Aird}, K.~A.,
  {Allam}, S.~S., {Armstrong}, R., {Barkhouse}, W.~A., {et~al.} 2010, arXiv
 1003.0005

\bibitem[{{High} {et~al.}(2009){High}, {Stubbs}, {Rest}, {Stalder}, \&
  {Challis}}]{High:2009zp}
{High}, F.~W., {Stubbs}, C.~W., {Rest}, A., {Stalder}, B., \& {Challis}, P.
  2009, \aj, 138, 110

\bibitem[{{Hoffleit}(1964)}]{Hoffleit:1964rt}
{Hoffleit}, D. 1964, {Catalogue of bright stars}, ed. {Hoffleit, D.}

\bibitem[{{H{\o}g} {et~al.}(2000){H{\o}g}, {Fabricius}, {Makarov}, {Urban},
  {Corbin}, {Wycoff}, {Bastian}, {Schwekendiek}, {et~al.}}]{Hog:2000ys}
{H{\o}g}, E., {Fabricius}, C., {Makarov}, V.~V., {Urban}, S., {Corbin}, T.,
  {Wycoff}, G., {Bastian}, U., {Schwekendiek}, P., {et~al.} 2000, \aap, 355,
  L27

\bibitem[{{Hsieh} {et~al.}(2005){Hsieh}, {Yee}, {Lin}, \&
  {Gladders}}]{Hsieh:2005fq}
{Hsieh}, B.~C., {Yee}, H.~K.~C., {Lin}, H., \& {Gladders}, M.~D. 2005, \apjs,
  158, 161

\bibitem[{{Ivezi{\'c}} {et~al.}(2004){Ivezi{\'c}}, {Lupton}, {Schlegel},
  {Boroski}, {Adelman-McCarthy}, {Yanny}, {Kent}, {Stoughton},
  {et~al.}}]{Ivezic:2004ad}
{Ivezi{\'c}}, {\v Z}., {Lupton}, R.~H., {Schlegel}, D., {Boroski}, B.,
  {Adelman-McCarthy}, J., {Yanny}, B., {Kent}, S., {Stoughton}, C., {et~al.}
  2004, Astronomische Nachrichten, 325, 583

\bibitem[{{Kepner} {et~al.}(1999){Kepner}, {Fan}, {Bahcall}, {Gunn}, {Lupton},
  \& {Xu}}]{kep}
{Kepner}, J., {Fan}, X., {Bahcall}, N., {Gunn}, J., {Lupton}, R., \& {Xu}, G.
  1999, \apj, 517, 78

\bibitem[{{Koester} {et~al.}(2007){Koester}, {McKay}, {Annis}, {Wechsler},
  {Evrard}, {Rozo}, {Bleem}, {Sheldon}, {et~al.}}]{Koester:2007ek}
{Koester}, B.~P., {McKay}, T.~A., {Annis}, J., {Wechsler}, R.~H., {Evrard},
  A.~E., {Rozo}, E., {Bleem}, L., {Sheldon}, E.~S., {et~al.} 2007, \apj, 660,
  221

\bibitem[{{Lawrence} {et~al.}(2007){Lawrence}, {Warren}, {Almaini}, {Edge},
  {Hambly}, {Jameson}, {Lucas}, {Casali}, {et~al.}}]{Lawrence:2007rp}
{Lawrence}, A., {Warren}, S.~J., {Almaini}, O., {Edge}, A.~C., {Hambly}, N.~C.,
  {Jameson}, R.~F., {Lucas}, P., {Casali}, M., {et~al.} 2007, \mnras, 379, 1599

\bibitem[{{Loh} {et~al.}(2008){Loh}, {Ellingson}, {Yee}, {Gilbank}, {Gladders},
  \& {Barrientos}}]{Loh:2007be}
{Loh}, Y., {Ellingson}, E., {Yee}, H.~K.~C., {Gilbank}, D.~G., {Gladders},
  M.~D., \& {Barrientos}, L.~F. 2008, \apj, 680, 214

\bibitem[{{Lonsdale} {et~al.}(2003){Lonsdale}, {Smith}, {Rowan-Robinson},
  {Surace}, {Shupe}, {Xu}, {Oliver}, {Padgett}, {et~al.}}]{Lonsdale:2003ix}
{Lonsdale}, C.~J., {Smith}, H.~E., {Rowan-Robinson}, M., {Surace}, J., {Shupe},
  D., {Xu}, C., {Oliver}, S., {Padgett}, D., {et~al.} 2003, \pasp, 115, 897

\bibitem[{{LSST Science Collaborations} {et~al.}(2009){LSST Science
  Collaborations}, {Abell}, {Allison}, {Anderson}, {Andrew}, {Angel}, {Armus},
  {Arnett}, {et~al.}}]{LSST-Science-Collaborations:2009uq}
{LSST Science Collaborations}, {Abell}, P.~A., {Allison}, J., {Anderson},
  S.~F., {Andrew}, J.~R., {Angel}, J.~R.~P., {Armus}, L., {Arnett}, D.,
  {et~al.} 2009, ArXiv e-prints

\bibitem[{{Lu} {et~al.}(2009){Lu}, {Gilbank}, {Balogh}, \&
  {Bognat}}]{Lu:2009gf}
{Lu}, T., {Gilbank}, D.~G., {Balogh}, M.~L., \& {Bognat}, A. 2009, \mnras, 399,
  1858

\bibitem[{{Majumdar} \& {Mohr}(2004)}]{Majumdar:2004nj}
{Majumdar}, S. \& {Mohr}, J.~J. 2004, \apj, 613, 41

\bibitem[{{Mitchell} {et~al.}(1976){Mitchell}, {Culhane}, {Davison}, \&
  {Ives}}]{mitchel76}
{Mitchell}, R.~J., {Culhane}, J.~L., {Davison}, P.~J.~N., \& {Ives}, J.~C.
  1976, \mnras, 175, 29P

\bibitem[{{Monet}(1998)}]{usno}
{Monet}, D.~G. 1998, American Astronomical Society Meeting, 30, 1427

\bibitem[{{Pettini} {et~al.}(2000){Pettini}, {Steidel}, {Adelberger},
  {Dickinson}, \& {Giavalisco}}]{Pettini:2000fk}
{Pettini}, M., {Steidel}, C.~C., {Adelberger}, K.~L., {Dickinson}, M., \&
  {Giavalisco}, M. 2000, \apj, 528, 96

\bibitem[{{Postman} {et~al.}(1996){Postman}, {Lubin}, {Gunn}, {Oke}, {Hoessel},
  {Schneider}, \& {Christensen}}]{pdcs}
{Postman}, M., {Lubin}, L.~M., {Gunn}, J.~E., {Oke}, J.~B., {Hoessel}, J.~G.,
  {Schneider}, D.~P., \& {Christensen}, J.~A. 1996, \aj, 111, 615+

\bibitem[{{Rasmussen} {et~al.}(2006){Rasmussen}, {Ponman}, {Mulchaey}, {Miles},
  \& {Raychaudhury}}]{Rasmussen:2006qy}
{Rasmussen}, J., {Ponman}, T.~J., {Mulchaey}, J.~S., {Miles}, T.~A., \&
  {Raychaudhury}, S. 2006, \mnras, 373, 653

\bibitem[{{Schlegel} {et~al.}(1998){Schlegel}, {Finkbeiner}, \&
  {Davis}}]{1998ApJ...500..525S}
{Schlegel}, D.~J., {Finkbeiner}, D.~P., \& {Davis}, M. 1998, \apj, 500, 525

\bibitem[{{Serlemitsos} {et~al.}(1977){Serlemitsos}, {Smith}, {Boldt}, {Holt},
  \& {Swank}}]{serlemitsos77}
{Serlemitsos}, P.~J., {Smith}, B.~W., {Boldt}, E.~A., {Holt}, S.~S., \&
  {Swank}, J.~H. 1977, \apjl, 211, L63

\bibitem[{{Treu} {et~al.}(2003){Treu}, {Ellis}, {Kneib}, {Dressler}, {Smail},
  {Czoske}, {Oemler}, \& {Natarajan}}]{Treu:2003ty}
{Treu}, T., {Ellis}, R.~S., {Kneib}, J., {Dressler}, A., {Smail}, I., {Czoske},
  O., {Oemler}, A., \& {Natarajan}, P. 2003, \apj, 591, 53

\bibitem[{{Visvanathan}(1978)}]{visv}
{Visvanathan}, N. 1978, \aap, 67, L17

\bibitem[{{Willott} {et~al.}(2005){Willott}, {Delfosse}, {Forveille},
  {Delorme}, \& {Gwyn}}]{Willott:2005eu}
{Willott}, C.~J., {Delfosse}, X., {Forveille}, T., {Delorme}, P., \& {Gwyn},
  S.~D.~J. 2005, \apj, 633, 630

\bibitem[{{Wuyts} {et~al.}(2010){Wuyts}, {Barrientos}, {Gladders}, {Sharon},
  {Bayliss}, {Carrasco}, {Gilbank}, {Yee}, {et~al.}}]{Wuyts:2010ul}
{Wuyts}, E., {Barrientos}, L.~F., {Gladders}, M.~D., {Sharon}, K., {Bayliss},
  M.~B., {Carrasco}, M., {Gilbank}, D.~G., {Yee}, H.~K.~C., {et~al.} 2010,
  \apj~submitted

\bibitem[{{Yee}(1991)}]{1991PASP..103..396Y}
{Yee}, H.~K.~C. 1991, \pasp, 103, 396

\bibitem[{{Yee} {et~al.}(1996){Yee}, {Ellingson}, \& {Carlberg}}]{Yee:1996lk}
{Yee}, H.~K.~C., {Ellingson}, E., \& {Carlberg}, R.~G. 1996, \apjs, 102, 269

\bibitem[{{Yee} {et~al.}(1998){Yee}, {Ellingson}, {Morris}, {Abraham}, \&
  {Carlberg}}]{Yee:1998rt}
{Yee}, H.~K.~C., {Ellingson}, E., {Morris}, S.~L., {Abraham}, R.~G., \&
  {Carlberg}, R.~G. 1998, \apjs, 116, 211

\end{thebibliography}

\end{document}